\RequirePackage[hyphens]{url}
\documentclass[acmsmall,screen]{acmart}
\usepackage{epsfig,subfigure,amsmath,amsfonts,graphicx,bm,color,multirow,stfloats,float,tabu,tabularx}
\usepackage{algorithm}
\usepackage{algorithmic}

\PassOptionsToPackage{hyphens}{url}

\AtBeginDocument{%
  }

\setcopyright{acmcopyright}
\copyrightyear{2025}
\acmYear{2025}
\acmDOI{10.1145/3722220}

\acmJournal{TOSN}
\acmVolume{21}
\acmNumber{5}
\acmArticle{3}
\acmMonth{5}

\begin{document}

\title{Enabling Low-Power Massive MIMO with Ternary ADCs for AIoT Sensing}

\author{Shengheng~Liu}
\authornote{Corresponding author}
\orcid{0000-0001-6579-9798}
\affiliation{%
  \institution{National Mobile Communications Research Laboratory, Southeast University}
  \city{Nanjing}
  \postcode{210096}
  \country{China}}
\affiliation{%
  \institution{Purple Mountain Laboratories}
  \city{Nanjing}
  \postcode{211111}
  \country{China}}
\email{s.liu@seu.edu.cn}

\author{Ningning~Fu}
\orcid{0009-0009-7407-6597}
\affiliation{%
  \institution{National Mobile Communications Research Laboratory, Southeast University}
  \city{Nanjing}
  \postcode{210096}
  \country{China}}
\email{funingning@seu.edu.cn}

\renewcommand{\shortauthors}{Liu et al.}

\begin{abstract}
The proliferation of networked devices and the surging demand for ubiquitous intelligence have given rise to the artificial intelligence of things (AIoT). However, the utilization of high-resolution analog-to-digital converters (ADCs) and numerous radio frequency chains significantly raises power consumption. This paper explores a cost-effective solution using ternary ADCs (T-ADCs) in massive multiple-input-multiple-output (MIMO) systems for low-power AIoT and specifically addresses channel sensing challenges. The channel is first estimated through a pilot-aided scheme and refined using a joint-pilot-and-data (JPD) approach. To assess the performance limits of this two-threshold ADC system, the analysis includes its hardware-ideal counterpart, the parallel one-bit ADCs (PO-ADCs) and a realistic scenario where noise variance is unknown at the receiver is considered. Analytical findings indicate that the JPD scheme effectively mitigates performance degradation in channel estimation due to coarse quantization effects under mild conditions, without necessitating additional pilot overhead. For deterministic and random channels, we propose modified expectation maximization (EM) and variational inference EM estimators, respectively. Extensive simulations validate the theoretical results and demonstrate the effectiveness of the proposed estimators in terms of mean square error and symbol error rate, which showcases the feasibility of implementing T-ADCs and the associated JPD scheme for greener AIoT smart sensing.
\end{abstract}




\received{6 February 2024}
%\accepted{3 March 2025}

\maketitle

\section{Introduction}

Mobile service providers rank among the largest energy consumers worldwide, responsible for up to three percent of the total global power usage. The growing increase in the contact-free sensing demands and densification of artificial intelligence of things (AIoT) facilities create substantial pressure for the service providers to adopt environmentally sustainable practices. This push is not only driven by ethical considerations but also by commercial imperatives. Given that the radio access network (RAN) accounts for the majority, $87\%$ as per \cite{gsma2023}, of telecom energy consumption, targeting RANs for green improvements is a logical priority \cite{cao2019stochastic}. Given that a significant proportion of AIoT devices are low-power battery-driven devices, considerable research interest has been directed towards the base station (BS) side. Massive multiple-input multiple-output (MIMO) technology \cite{you2023toward, yan2022opencarrier, nguyen2014cooperative} is capable of creating a large number of spatial degrees of freedoms, which are essential in achieving high spectral efficiency, high data rates, and spatial multiplexing. In typical massive MIMO setups, hundreds of antennas are employed, each connected to a dedicated radio frequency (RF) chain equipped with high-resolution (e.g., 12–-16 bits) analog-to-digital converters (ADCs) \cite{buiquang2019blind}. However, ADCs are known to consume a substantial amount of power dissipated at a base station receiver, and the power consumption grows exponentially with resolution level and linearly with sampling rate \cite{rangan2014millimeter}. Consequently, high-resolution quantization poses a significant challenge to the energy efficiency of massive MIMO based AIoT systems is inviable commercially.

While low-resolution ADC deployments enable economically attractive receivers, coarse quantization inevitably leads to a fundamental shift, resulting in highly distorted and nonlinear quantized signals. Consequently, using conventional linear estimators/detectors to handle quantized signals undergoes inherent performance degeneration \cite{choi2016near,saxena2017analysis}, especially in one-bit quantization systems. Channel estimators for quantized MIMO systems are mainly developed under the principle of maximum likelihood (ML)  \cite{choi2016near,mezghani2018blind} or Bayes rules \cite{wen2015bayes, ding2018bayesian, zhang2017one} in the literature. In \cite{li2017channel}, a Bussgang decomposition based channel estimator for one-bit massive MIMO systems was developed. In \cite{mezghani2010multiple}, the ML estimates of multiple parameters of interest within quantized observations were obtained by the expectation maximization (EM) algorithm.  In \cite{wen2015bayes}, the generalized approximate message passing based channel estimator was proposed for the quantized massive MIMO system, which has been proven to attain the optimal Bayes limit asymptotically, and its enhanced variant, namely vector approximate message passing algorithm, to deal with the broadband millimeter channel, was investigated in \cite{mo2017channel}. In \cite{ding2018bayesian}, a hybrid analog-digital preprocessing structure was integrated with the quantized mmWave massive MIMO system, which develops a super-resolution beamspace channel estimator under the sparse Bayesian learning framework. A more advanced antithetic dithered one-bit receiver was analyzed in \cite{ho2019antithetic}, and the proposed parameter expanded EM channel estimator has been verified to achieve fast convergence and monotonic MSE performance over a wide signal-to-noise ratio (SNR) range.

Throughput analyses in works such as \cite{singh2009limits, mo2015capacity, jacobsson2017throughput} and error bound studies in \cite{li2017channel, rao2019channel, liu2020angular} reveal a nonnegligible and systematic performance loss in one-bit systems compared to their full-resolution counterparts. To mitigate this performance degradation without incurring substantial hardware overhead, the ternary ADC (T-ADC) architecture was developed in \cite{tanoue2007ternary, zhang2020massive}. Unlike conventional integer-bit ADC, the T-ADC compares the analog input against two nonidentical thresholds, generating quantized outputs in three possible intervals. In massive MIMO systems, T-ADCs offer a middle ground in power consumption, sitting between one-bit and two-bit ADCs, yet surpass both in energy efficiency \cite{tanoue2007ternary, zhang2020massive}. From a physical perspective, as a type of two-threshold ADC, the hardware-ideal counterpart of the T-ADC is the parallel one-bit ADC (PO-ADC), as depicted in Fig. \ref{Fig:1}.
This paper explores the challenge of channel estimation using low-resolution ADCs in AIoT sensing applications and provides theoretical error bounds concerning various system parameters. Additionally, we propose two channel estimators, one tailored for deterministic channels and the other for random channels, to boost transmission performance in massive MIMO systems for AIoT sensing. The technical contributions of this paper are summarized as follows.
\begin{itemize}
	\item We consider the situation where the noise variance is unknown at the receiver \cite{liu2020angular, ribeiro2006bandwidth}. For pilot-aided (PA) channel estimation, we derive the Cram{\'e}r-Rao lower bounds (CRLBs) on the channel parameter estimates to determine the performance limit of such a quantized massive MIMO system. We also deduce analytically expressible CRLBs for the single-input-single-output (SISO) system, which explicitly describe how the relative distance between the ADCs' thresholds and the channel magnitude affects the CRLBs. Analytical and numerical results show that the T-ADC achieves comparable error performance to the PO-ADC over a wide SNR range, with reduced power consumption.
	
	\item We propose a joint-pilot-and-data (JPD) scheme to leverage all the transmitted symbols in channel estimation, significantly reducing pilot overhead, which can be prohibitive for massive MIMO systems. Due to the coarse quantization of the T-ADC, the robustness of the PA estimation heavily relies on the pilot length. Rigorous CRLB analysis shows that the JPD scheme is comparable to the PA one under mild conditions. To the best of our knowledge, this is the first attempt in theory to reveal its effectiveness in quantized MIMO systems.
	
	\item We develop channel estimators under the principle of expectation maximization (EM). For the deterministic channel, a modified EM method is proposed to provide ML estimates and its parallel implementations are further designed to lower the computational complexity with increasing  number of antennas and observation length. For the random channel, a variational inference EM (VIEM) estimator is developed by solving the quantized sparse Bayesian learning problem.
\end{itemize}

\textit{Notation:} The lowercase letter $x$, bold lowercase letter $\bf{x}$, and bold uppercase letter  $\bf{X}$ denote scalar quantity, column vector and matrix, respectively. The symbols ${[ {\bf{X}} ]_{k,n}}$, ${[ {\bf{X}} ]_{k,:}}$ and ${[ {\bf{X}} ]_{:,n}}$ respectively denote the $k$-th row and $n$-th column element, the $k$-th row, and the $n$-th column, of $\bf{X}$. The superscripts ${\left(  \cdot  \right)^{\mathsf T}}$ and ${\left(  \cdot  \right)^{\mathsf H}}$ denote transpose and conjugate transpose, respectively. The operators $\rm{Re}(\cdot)$ and $\rm{Im}(\cdot)$ extract respectively the real part and the imaginary part  of a complex-valued quantity. The function ${\mathcal{Q}}( \cdot )$ refers to the quantization process, $\Pr \left(  \cdot  \right)$ denotes the probability, ${{\mathbb{E}}}\left[  \cdot  \right]$ denotes the statistical expectation, {${\mathcal{O}}(\cdot)$ denotes the complexity}, and $\mathop  = \limits^{{\rm{a}}{\rm{.e}}{\rm{.}}} $ means an asymptotic equality. The operator ${\rm{vec(}} {\bf X} {\rm{)}}$ vectorizes a matrix ${\bf X}$, $\rm{tr}({\bf X})$ denotes the trace of ${\bf X}$, and ${\rm{diag(}}{\bf{X}}{\rm{)}}$ denotes a column vector composed of the diagonal elements of $\bf X$. Given an $N\times 1$ vector ${\bf{x}} = \left[ {{x_1}, {{x_2}}, \cdots, {x_N}} \right]^{\mathsf T}$ and a real-valued continuous and differentiable function $f(\bf{x})$, the operator ${\nabla _{\bf{x}}}{f} \left( {\bf{x}} \right) = {\left[ {\partial f \left( {\bf{x}} \right)/\partial {x_1},  \partial f\left( {\bf{x}} \right)/\partial {x_2}, \cdots, \partial f \left( {\bf{x}} \right)/\partial {x_N}} \right]^{\mathsf T}}$ denotes the gradient, and the operator ${\rm{diag(}}{\bf{x}}{\rm{)}}$ returns an $N\times N$ diagonal matrix with $ \left\{ {{x_1}, {{x_2}}, \cdots, {x_N}} \right\}$ being its diagonal elements.

\section{System Model}
\label{sec:sys}

\begin{figure}[t]
	%	\vspace{-1cm}
	\hspace{-2cm}
	\centering
	{\begin{minipage}[h]{0.5\textwidth}
			\centering
			\epsfig{figure=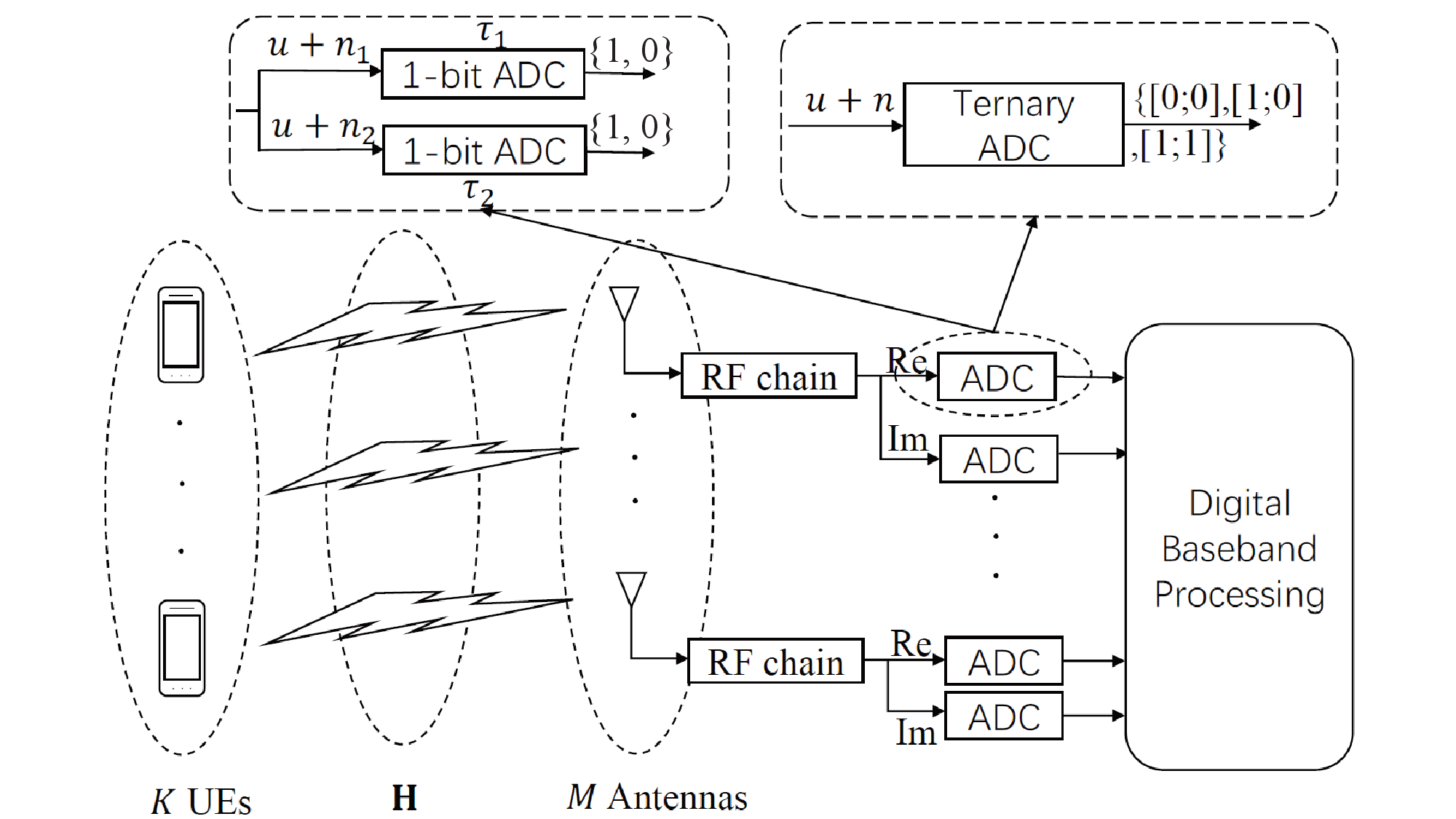,width=1.3\columnwidth}
		\end{minipage}
	}
	\hfill
	\caption{Diagram of uplink massive MIMO system with two-threshold ADCs.}	
	\label{Fig:1}
\end{figure}

Consider an uplink single-cell massive MIMO system with $K$ single-antenna AIoT user equipments (UEs) and an $M$-antenna BS, where $M \gg K$, as depicted in Fig. \ref{Fig:1}. We assume the channel is narrowband and Rayleigh block fading, i.e., the channel remains constant over $T$ consecutive intervals, and $T \ge K$. When all the $K$ UEs simultaneously transmit independent symbols to the BS with the same power $P_{\rm{s}}$, the received signal ${\bf{Y}} \in {\mathbb{C}^{M \times T}}$ at the BS  can be expressed as
\begin{eqnarray}
	{\bf{Y}} = {\bf{HX}} + {\bf{N}}, \label{eq:sys_mod}
\end{eqnarray}
where ${\bf{X}} \in {\mathbb{C}^{K \times T}}$ is the transmitted symbol matrix, and ${\mathbb{E}}[{[{\bf{X}}]_{k,t}}] = 0$ and ${\mathbb{E}}[|[{\bf{X}}]_{k,t}|] = {\sqrt{P_{\rm{s}}}}$. The channel matrix ${\bf{H}} \in {\mathbb{C}^{M \times K}}$ contains the fading coefficients between the transmitting antennas and the receiving ones, which can be deterministic or random. ${\bf{N}} \in {\mathbb{C}^{M \times T}}$ is the additive temporally and spatially white Gaussian noise with zero mean and element-wise variance $\sigma^2$.

At the BS, the received signal is first down converted into the real/imaginary analog baseband. Then, a variable gain amplifier with an automatic gain control is implemented at the front end to ensure that the amplitude of analog baseband signal is within a proper range for subsequent analog-to-digital process \cite{wen2015bayes}. We consider that the signal is quantized by the T-ADC and its counterpart PO-ADC, whose architectures are described as follows.

\textit{Ternary ADC (T-ADC) \cite{tanoue2007ternary}: } {Suppose that a real-valued analog signal $u$, corrupted by a noise $n$, is quantized by a T-ADC with two nonidentical thresholds $\tau_1 = -\tau_2<0$, the quantized output ${\bf{z}}^{\mathsf T} \in {\mathbb{R}}^{1\times2}$ is given as}
\begin{eqnarray}
{\bf{z}}^{\mathsf T} = {\mathop{\mathcal Q}}_{\rm{t}} ([u+n,u+n]) = \left\{ {\begin{array}{*{20}{c}}
	{{{[0,0]}}},&{u+n < {\tau _1}};\\
	{{{[1,0]}}},&{{\tau _1} \le u+n < {\tau _2}};\\
	{{{[1,1]}}},&{u+n \ge {\tau _2}}.
	\end{array}} \right.
\label{eq:tern}
\end{eqnarray}

According to \eqref{eq:tern}, the T-ADC is a kind of two-threshold ADCs, because its analog input is compared with two nonidentical thresholds $\tau_1$ and $\tau_2$. A similar hardware implementation considers a parallel-one-bit ADC, which is detailed as follows.

\textit{Parallel-one-bit ADC (PO-ADC) \cite{ho2019antithetic}:} The real-valued analog signal $u$, corrupted by uncorrelated noises $n_i,i=1,2$, is separately quantized by two parallelly connected one-bit ADCs with the threshold pair $\tau_1=-\tau_2<0$.  The $i$-th element of the quantized output ${\bf{z}}^{\mathsf T} \in {\mathbb{R}}^{1\times2}$ is given by
\begin{eqnarray}
{[{{\bf{z}}^{\mathsf T}]}_i}{\mathcal{ = Q}}_{\rm{p}}(u + {n_i}) = \left\{ {\begin{array}{*{20}{c}}
	0,&{{\rm{}}{\kern 1pt} {\kern 1pt} {\kern 1pt} {\kern 1pt} u + {n_i} < {\tau _i}};\\
	{  1},&{{\rm{}}{\kern 1pt} {\kern 1pt} {\kern 1pt} {\kern 1pt} u + {n_i} \ge {\tau _i}}.
	\end{array}} \right.
\label{eq:poadc}
\end{eqnarray}

In practice, the noises $n_i,i=1,2,$ from two analog branches of PO-ADC are positively correlated in the bandlimited systems. To provide physical insights into such two-threshold ADC systems, we consider a simplification that the noises $n_i$, $i=1,2$, are independent \cite{ho2019antithetic}. By comparing \eqref{eq:tern} and \eqref{eq:poadc}, for the noise corrupted signal, the PO-ADC generates more possible quantized outputs than the T-ADC, {i.e., its output can be ${\bf{z}} = [0,1]^{\mathsf T}$, which is inexistent in the T-ADC. This intuitively indicates that the error performance of channel estimation with PO-ADCs should be better than that with T-ADCs, because of additional data measurements.}

For ease of analysis, we represent the complex-value signal in a bivariate real form. By denoting ${\bf{S}} = {\bf{HX}} = {[{{\bf{s}}_1},{{\bf{s}}_2}, \ldots ,{{\bf{s}}_M}]^{\mathsf T}}$ and ${\bf{g}}_m \in {\mathbb{C}^{K \times 1}}$ as the channel vector between $K$ UEs and the $m$-th receiving antenna, i.e., ${{\bf{H}}} = [{{\bf{g}}_1},{{\bf{g}}_2}, \ldots ,{{\bf{g}}_M}]^{\mathsf T}$, the bivariate real representation of ${{\bf{s}}_m}\in {\mathbb{R}^{T \times 1}}$ is given by
\begin{eqnarray}
{{{\bf{\bar s}}}_m} = \left[ {\begin{array}{*{20}{c}}
	{{\mathop{\rm Re}\nolimits} ({{\bf{s}}_m})}\\
	{{\mathop{\rm Im}\nolimits} ({{\bf{s}}_m})}
	\end{array}} \right]  = {\bf{\bar X}}{{{\bf{\bar g}}}_m}   \in {\mathbb{R}^{2T \times 1}},
\label{eq:RI}
\end{eqnarray}
where
\[{\bf{\bar X}} = \left[ {\begin{array}{*{20}{c}}
	{{\mathop{\rm Re}\nolimits} ({{\bf{X}}^{\mathsf T}})}&{ - {\mathop{\rm Im}\nolimits} ({{\bf{X}}^{\mathsf T}})}\\
	{{\mathop{\rm Im}\nolimits} ({{\bf{X}}^{\mathsf T}})}&{{\mathop{\rm Re}\nolimits} ({{\bf{X}}^{\mathsf T}})}
	\end{array}} \right]  \in {\mathbb{R}^{2T \times 2K}},{{{\bf{\bar g}}}_m} = \left[ {\begin{array}{*{20}{c}}
	{{\mathop{\rm Re}\nolimits} ({{\bf{g}}_m})}\\
	{{\mathop{\rm Im}\nolimits} ({{\bf{g}}_m})}
	\end{array}} \right]  \in {\mathbb{R}^{2K \times 1}}.\]
In this way, the quantized output becomes
\begin{eqnarray}
	{\bf{\bar Z}}_m = {\mathop{\mathcal Q}}_{\rm{p/t}} ( [{{{\bf{\bar s}}}_m},  {{{\bf{\bar s}}}_m}]  + {{\bf{\bar N}}} )\in {\mathbb{R}^{2T \times 2}},
	\label{eq:Z}
\end{eqnarray}
where  the quantization ${\mathop{\mathcal Q}}_{\rm{p/t}}(\cdot)$ is applied component-wise, according to {\eqref{eq:tern} or \eqref{eq:poadc},} and ${{\bf{\bar N}}}\in {\mathbb{R}^{2T \times 2}}$ is the additive white Gaussian noise with zero mean and element-wise variance $\sigma^2/2$. In the PO-ADC, $[{{\bf{\bar N}}}]_{i,1}$ and $[{{\bf{\bar N}}}]_{i,2}$, $\forall i$ are uncorrelated, and in the T-ADC, we have $[{{\bf{\bar N}}}]_{i,1}=[{{\bf{\bar N}}}]_{i,2},\forall i$.

\section{Pilot-Aided Channel Estimation}
\label{sec:PA}

In this section, we consider the pilot-aided (PA) channel estimation, where the {length-$T_{\rm{p}}$ pilot matrix ${\bf{X}} \in {\mathbb{C}^{K \times T_{\rm{p}}}}$ and its bivariate real form ${\bf{\bar X}} \in {\mathbb{R}^{2T_{\rm{p}}\times 2K} }$} are known but the deterministic noise variance $\sigma^2$ is unknown at the BS. {To evaluate the error performance in the two-threshold ADC system, the CRLBs for both deterministic and random channels are studied.} As discussed in \cite{choi2016near,ding2018bayesian,liu2020angular}, the robustness of channel estimation in a quantized system heavily relies on how accurate the acquisition of the noise variance is. Since the channel vector ${\bf{\bar g}}_m$ can be successfully estimated from the quantized output ${\bf{\bar Z}}_m$, in the sequel, whenever it is clear from the context, the antenna index $m$ has been omitted for mathematical elegance, and the parameter set of interest is defined as ${\bm{\eta }} = {[{{\bf{\bar g}}^{\mathsf T}},\sigma ]^{\mathsf T}} \in {\mathbb{R}^{(2K+1) \times 1}}$.

\subsection{Deterministic CRLB and ML Estimation}
\label{sec:dete}

When the channel ${\bf{\bar g}}$ is deterministic but unknown at the BS, the MSE of the channel estimate $ {\bf \hat{\bar g}}$ is given by
\begin{eqnarray}
{\rm{mse}}( { { {\bf \hat{\bar g}}}} ) = {{\mathbb{E}}}{_{{ { {\bf \hat{\bar g}}}}|{\bf{\bar g}}}}\left[ {\left\| {{ { {\bf \hat{\bar g}}}} - {\bf{\bar g}}} \right\|_2^2} \right].
\label{eq:MSEB}
\end{eqnarray}
The MSE of any unbiased and asymptotically efficient estimator reaches the deterministic CRLB as the {pilot length $T_{\rm{p}}$} increases \cite{kay1993fundamentals}, which is detailed as follows.

\subsubsection{PO-ADC Systems}
{For the pilot matrix ${\bf{\bar X}} \in {\mathbb{R}^{2T_{\rm{p}}\times 2K} }$ and the associated quantized output ${\bf{\bar Z}} \in {\mathbb{R}^{2T_{\rm{p}} \times 2}}$,} the log-likelihood-function (LLF) of the PO-ADC system is given by
\begin{eqnarray}
	{L_{\rm{p}}}( {{\bf{\bar Z}};{\bm{\eta }}} )= {\rm{ln}} {{\rm{Pr(}}{\bf{\bar Z}},{\bm{\eta }}{\rm{)}}}  = \sum\limits_{i = 1}^{2{T_{\rm{p}}}} {\sum\limits_{j = 1}^2 {\ln Q\!\left( {\left({{2[{\bf{\bar Z}}]}_{i,j}}-1\right){{[{{\bf{A}}]_{i,j}}}}} \right)} },
	\label{eq:likli}
\end{eqnarray}
where  ${{Q}}( x ) = \left(1/\sqrt {2\uppi } \right)\int_{ - \infty }^x {{e^{ - {u^2}/2}}du}  = \int_{ - \infty }^x {{{q}}(u)du} $ is the cumulative distribution function (CDF) of the standard Gaussian probability density function (PDF)  ${{q}}(x)$, and
\begin{eqnarray}
	{[{{\bf{A}}}]_{i,j}} = \frac{{\sqrt 2 \left( {{{[{{\bf{\bar X}}}{\bf{\bar g}}]}_i} - {\tau _j}} \right)}}{\sigma },i = 1,2, \ldots ,2{T_{\rm{p}}},{\kern 1pt} {\kern 1pt} {\kern 1pt} {\kern 1pt} j = 1,2.
	\label{eq:dA}
\end{eqnarray}

The Fisher information matrix (FIM) ${{\bf{F}}_{\rm{p}}}({\bm{\eta }})\in {\mathbb{R}^{(2K+1) \times (2K+1)}}$ is given by \cite{kay1993fundamentals}
\begin{eqnarray}
	{{\bf{F}}_{\rm{p}}}({\bm{\eta }}) = {{\mathbb{E}}_{{\bf{\bar Z}}|{\bm{\eta }}}}\left[ {{\nabla _{\bm{\eta }}}{{L_{\rm p}}}( {{\bf{\bar Z}};{\bm{\eta }}} ){{\left( {{\nabla _{\bm{\eta }}}{{L_{\rm p}}}( {{\bf{\bar Z}};{\bm{\eta }}})} \right)}^{\mathsf T}}} \right],
	\label{eq:FIM}
\end{eqnarray}
and the deterministic CRLB is the inverse of the FIM ${{\bf{F}}_{\rm{p}}}({\bm{\eta }})$, i.e.,
\begin{eqnarray}
	{\bf{CRLB}} ( {\bm{\eta }} ){\rm{ = }}{\left( {{{\bf{F}}_{\rm{p}}}}\left({\bm{\eta }}\right) \right)^{ - 1}},
	\label{eq:CRB}
\end{eqnarray}
whose analytical expression is discussed in the following \textit{Theorem 1}.

\textit{Theorem 1:} In the PO-ADC system, the FIM for the PA channel estimation is ${\bf{F}}_{\rm{p}}( {\bm{\eta }} ) = {\bf{M\Lambda }}{{\bf{M}}^{\mathsf T}}$, where ${\bf{M}}$ and ${\bf{\Lambda }}$ are given in \eqref{eq:fiser1}. The MSE of the ML estimate ${ { {\bf \hat{\bar g}}}}$ can be expressed as
\begin{eqnarray}
	{\rm{mse_{PA,p}}}({ { {\bf \hat{\bar g}}}}) \mathop  = \limits^{{\rm{a}}{\rm{.e}}{\rm{.}}} {\rm{tr}}\left( {{{\left[ {{{\left( {{\bf{M\Lambda }}{{\bf{M}}^{\mathsf T}}} \right)}^{ - 1}}} \right]}_{1:2K,1:2K}}} \right).
	\label {eq:mse1}
\end{eqnarray}

\textit{Proof:} See Appendix A.   \hfill $\blacksquare$\\

The FIM ${{\bf{F}}_{\rm{p}}}({\bm{\eta }})$ derived in \textit{Theorem 1} can be rewritten in a block matrix form as
\begin{eqnarray}
	{\bf{F}}_{\rm{p}}({\bm{\eta}}) ={\bf{M\Lambda }}{{\bf{M}}^{\mathsf T}}= \left[ {\begin{array}{*{20}{c}}
		{{\bf{F}}_{\rm{p}}({\bf{\bar g}})}&{{\bf{f}}_{\rm{p}}({\bf{\bar g}},\sigma )}\\
		{{{\left( {{\bf{f}}_{\rm{p}}({\bf{\bar g}},\sigma )} \right)}^{\mathsf T}}}&{{ f_{\rm p}}(\sigma )}
		\end{array}} \right],
	\label{eq:blockF}
\end{eqnarray}
where ${{\bf{F}}_{\rm{p}}}({\bf{\bar g}})$ follows the structure of ${\bf{F}}_{\rm{p}}({\bm{\eta}})$ in \eqref{eq:fiser} by substituting the term ${\partial {{[{{\bf{A}}}]}_{i,j}}/\partial {\bm{\eta}}}$  with ${\partial {{[{{\bf{A}}}]}_{i,j}}/\partial {\bf{\bar g}}}$, so that ${{\bf{F}}_{\rm{p}}}({\bf{\bar g}}) = {{{[{\bf{M}}]}_{1:2K,:}}{\bf{\Lambda }}{{[{{\bf{M}}^{\mathsf T}}]}_{:,1:2K}}}$,
\begin{eqnarray*}
	{\bf{f}}_{\rm{p}}({\bf{\bar g}},\sigma ) \!\!\!\!\!&=&\!\!\!\!\! \sum\limits_{i = 1}^{{2T_{\rm{p}}}} {\sum\limits_{j = 1}^2 {\frac{{{{{q}}^2}{\rm{([}}{{\bf{A}}}{{\rm{]}}_{i,j}}{\rm{)}}}}{{{{Q([}}{{\bf{A}}}{{\rm{]}}_{i,j}}{{)Q(}} - {{{\rm{[}}{{\bf{A}}}{\rm{]}}}_{i,j}}{\rm{)}}}}\frac{{\partial {{{\rm{[}}{{\bf{A}}}{\rm{]}}}_{i,j}}}}{{\partial {\bf{\bar g}}}}} {{ {\frac{{\partial {{{\rm{[}}{{\bf{A}}}{\rm{]}}}_{i,j}}}}{{\partial \sigma }}} }}},
\end{eqnarray*}
and
\begin{eqnarray*}
{ f}_{\rm{p}}(\sigma ) \!\!\!\!\!&=&\!\!\!\!\! {\sum\limits_{i = 1}^{{2T_{\rm{p}}}} {\sum\limits_{j = 1}^2 {\frac{{{{{q}}^2}{\rm{([}}{{\bf{A}}}{{\rm{]}}_{i,j}}{\rm{)}}}}{{{{Q([}}{{\bf{A}}}{{\rm{]}}_{i,j}}{{)Q(}} - {{{\rm{[}}{{\bf{A}}}{\rm{]}}}_{i,j}}{\rm{)}}}}\left( {\frac{{\partial {{{\rm{[}}{{\bf{A}}}{\rm{]}}}_{i,j}}}}{{\partial \sigma }}} \right)} } ^2}.
\end{eqnarray*}

Then the MSE of ${ { {\bf \hat{\bar g}}}}$ in \eqref{eq:mse1} can be expressed in a more compact form as
\begin{eqnarray}
{\rm{mse_{PA,p}}}\left({ { {\bf \hat{\bar g}}}}\right) \!\!\!\!\!&\mathop  = \limits^{{\rm{a}}{\rm{.e}}{\rm{.}}}&\!\!\!\!\! {\rm{tr}}\!\left( {{{\left[ {{{\left( {{\bf{F}}_{\rm{p}}({\bm{\eta}})} \right)}^{ - 1}}} \right]}_{1:2K,1:2K}}} \right)
\mathop {\rm{ = }}\limits^{({\rm{a}})} {\rm{tr}}\!\left( {{{\left( {{\bf{F}}_{\rm{p}}({\bf{\bar g}}) \!-\! {\bf{f}}_{\rm{p}}({\bf{\bar g}},\sigma ){{\left( {{f_{\rm p}}(\sigma )} \right)}^{ - 1}}{{\left( {{\bf{f}}_{\rm{p}}({\bf{\bar g}},\sigma )} \right)}^{\mathsf T}}} \right)}^{ - 1}}}\! \right) \nonumber \\
\!\!\!\!\!&\mathop {\rm{ = }}\limits^{({\rm{b}})} &\!\!\!\!\! {\rm{tr}}\!\left( {{{\left( {{\bf{F}}_{\rm{p}}({\bf{\bar g}})} \right)}^{ - 1}} + {\bf{B}}} \right),
\label{eq:block}
\end{eqnarray}
where \[{\bf{B}} = \frac{{{{\left( {{f_{\rm p}}(\sigma )} \right)}^{ - 1}}{{\left( {{\bf{F}}_{\rm{p}}({\bf{\bar g}})} \right)}^{ - 1}}{\bf{f}}_{\rm{p}}({\bf{\bar g}},\sigma ){{\left( {{\bf{f}}_{\rm{p}}({\bf{\bar g}},\sigma )} \right)}^{\mathsf T}}{{\left( {{\bf{F}}_{\rm{p}}({\bf{\bar g}})} \right)}^{ - 1}}}}{{1 - {{\left( {{f_{\rm p}}(\sigma )} \right)}^{ - 1}}{{\left( {{\bf{f}}_{\rm{p}}({\bf{\bar g}},\sigma )} \right)}^{\mathsf T}}{{\left( {{\bf{F}}_{\rm{p}}({\bf{\bar g}})} \right)}^{ - 1}}{\bf{f}}_{\rm{p}}({\bf{\bar g}},\sigma )}},\] and (a) and (b) stem from the block matrix inversion and the Sherman-Morrison lemma \cite{horn2012matrix}, respectively. The following \textit{Theorem 2} provides more analytical insights into the PO-ADC system.

\textit{Theorem 2:} Suppose that the noise variance $\sigma^2$ is known at the BS and ${\bf{X}}{{\bf{X}}^{\mathsf H}} = {T_{\rm{p}}}{P_{\rm{s}}}{{\bf{I}}_K}$ (orthogonal pilot), the minimum MSE of the ML channel estimate ${\bf \hat{\bar g}}$ in the PO-ADC system is
\begin{eqnarray}
	\min {\rm{ms}}{{\rm{e}}_{{\rm{PA}},{\rm{p}}}}( {{\bf\hat{ \bar g}}} ) \ge \frac{{K\uppi {\sigma ^2}}}{{4{P_{\rm{s}}}{T_{\rm{p}}}}}.
\end{eqnarray}
Moreover, by denoting ${\rm{mse}}{_{{\rm{f}}}}({\bf \hat{\bar g}})$ as the MSE of the ML channel estimate  in the full resolution ADC system, we have
\begin{eqnarray}
	 {\rm{ms}}{{\rm{e}}_{{\rm{PA}},{\rm{p}}}}( {{\bf\hat{ \bar g}}} ) \ge	\frac{\uppi}{2 }{\rm{ms}}{{\rm{e}}_{\rm{f}}}( {{\bf\hat{ \bar g}}} ).
	\label{eq:lower_mse1}
\end{eqnarray}

{\textit{Proof:} See Appendix B.   \hfill $\blacksquare$\\}

Equation \eqref{eq:lower_mse1} indicates that at least, there exists a $10 \textmd{log}_{\textmd{10}}(2/\uppi)\approx 1.96$ dB performance loss in the ML channel estimate in the PO-ADC system,  compared with its full resolution counterpart.

\subsubsection{T-ADC Systems}

The above analysis is focused on the hardware-ideal PO-ADC to characterize the performance limit of such two-threshold ADC systems. We next derive the deterministic CRLB of the T-ADC system. Invoking the quantization rule in \eqref{eq:tern}, {for the pilot matrix ${\bf{\bar X}} \in {\mathbb{R}^{2T_{\rm{p}}\times 2K} }$ and the associated quantized output ${\bf{\bar Z}} \in {\mathbb{R}^{2T_{\rm{p}} \times 2}}$,} the LLF becomes
\begin{eqnarray}
{L_{\rm{t}}}( {{\bf{\bar Z}};{\bm{\eta }}} )\!=\!\!\sum\limits_{i = 1}^{2{T_{\rm{p}}}}\! \left(\left( {\!1\!\!-\!\!{{[{\bf{\bar Z}}]}_{i,1}\!}} \right)\ln\!Q\!\left( {-{{[{\bf{A}}]}_{i,1}}} \right) \!\!+\!\! \left( {{{[{\bf{\bar Z}}]}_{i,1}}\!\!-\!\!{{[{\bf{\bar Z}}]}_{i,2}}} \right)
\!\cdot\!{\ln}\!\!\left( {Q\!\!\left( {{{[{\bf{A}}]}_{i,1}}} \right) \!-\! Q\!\!\left( {{{[{\bf{A}}]}_{i,2}}} \right)} \right) \!\!+\!\! {[{\bf{\bar Z}}]_{i,2}}\!\ln\!Q\!\left( {{{[{\bf{A}}]}_{i,2}}} \right)\right).
\label{eq:llft}
\end{eqnarray}

Based on \eqref{eq:FIM} and \eqref{eq:CRB}, the deterministic CRLB is given in \textit{Theorem 3}.

\textit{Theorem 3:}
In the T-ADC system, the FIM for the PA channel estimation ${{\bf{F}}_{\rm{t}}}({\bm{\eta }})$ is given by
\begin{eqnarray}
{{\bf{F}}_{\rm{t}}}({\bm{\eta }}) \!\!\!\!\!&=&\!\!\!\!\! \sum\limits_{i = 1}^{2{T_{\rm{p}}}} \Bigg[\left( {\frac{{{q^2}({{[{\bf{A}}]}_{i,1}})}}{{Q({{[{\bf{A}}]}_{i,1}})}} + \frac{{{q^2}({{[{\bf{A}}]}_{i,1}})}}{{Q({{[{\bf{A}}]}_{i,2}}) - Q({{[{\bf{A}}]}_{i,1}})}}} \right)\frac{{\partial {{[{\bf{A}}]}_{i,1}}}}{{\partial {\bm{\eta }}}}{{\left( {\frac{{\partial {{[{\bf{A}}]}_{i,1}}}}{{\partial {\bm{\eta }}}}} \right)}^{\mathsf T}} \nonumber \\ \!\!\!\!\!&-&\!\!\!\!\! \frac{{q({{[{\bf{A}}]}_{i,1}})q({{[{\bf{A}}]}_{i,2}})}}{{Q({{[{\bf{A}}]}_{i,2}}) - Q({{[{\bf{A}}]}_{i,1}})}}\frac{{\partial {{[{\bf{A}}]}_{i,1}}}}{{\partial {\bm{\eta }}}}{{\left( {\frac{{\partial {{[{\bf{A}}]}_{i,2}}}}{{\partial {\bm{\eta }}}}} \right)}^{\mathsf T}} - \frac{{q({{[{\bf{A}}]}_{i,1}})q({{[{\bf{A}}]}_{i,2}})}}{{Q({{[{\bf{A}}]}_{i,2}}) - Q({{[{\bf{A}}]}_{i,1}})}}\frac{{\partial {{[{\bf{A}}]}_{i,2}}}}{{\partial {\bm{\eta }}}}{\left( {\frac{{\partial {{[{\bf{A}}]}_{i,1}}}}{{\partial {\bm{\eta }}}}} \right)^{\mathsf T}} \nonumber\\
\!\!\!\!\!&+&\!\!\!\!\! \left( {\frac{{{q^2}({{[{\bf{A}}]}_{i,2}})}}{{Q({{[{\bf{A}}]}_{i,2}}) - Q({{[{\bf{A}}]}_{i,1}})}} + \frac{{{q^2}({{[{\bf{A}}]}_{i,2}})}}{{Q({{-[{\bf{A}}]}_{i,2}})}}} \right)\frac{{\partial {{[{\bf{A}}]}_{i,2}}}}{{\partial {\bm{\eta }}}}{\left( {\frac{{\partial {{[{\bf{A}}]}_{i,2}}}}{{\partial {\bm{\eta }}}}} \right)^{\mathsf T}}\Bigg].
\label{eq:fimt}
%\tag{27}
\end{eqnarray}
The MSE of the ML estimate ${\bf \hat{\bar g}}$ can be expressed as
\begin{eqnarray}
{\rm{mse_{PA,t}}}({\bf \hat{\bar g}}) \mathop  = \limits^{{\rm{a}}{\rm{.e}}{\rm{.}}} {\rm{tr}}\left( {{{\left[ {{{\left( {{\bf{F}}_{\rm{t}}}({\bm{\eta}}) \right)}^{ - 1}}} \right]}_{1:2K,1:2K}}} \right).
\label {eq:mse2}
\end{eqnarray}

\textit{Proof:} The proof is similar to that of \textit{Theorem 1} in Appendix A, and we omit the details for simplicity.

Although the FIM ${{\bf{F}}_{\rm{t}}}({\bm{\eta }})$ in \eqref{eq:fimt} can be also represented in a block matrix form as ${{\bf{F}}_{\rm{p}}}({\bm{\eta }})$ in \eqref{eq:blockF},  a further investigation on its inverse will not provide any insight into the performance limit of the  two-threshold  ADC system, due to the coupling nature between ${{{[{\bf{A}}]}_{i,1}}}$ and ${{{[{\bf{A}}]}_{i,2}}}$ in \eqref{eq:fimt}. In fact, the CRLB of $\bm{\eta }$ in the T-ADC system is sufficiently close to that in the PO-ADC system over a wide SNR range, which will be illustrated by the numerical results in Section \ref{sec:simu}. The above analysis only pertains to the $m$-th antenna. Given $\mathbf{H} = [\mathbf{g}_1, \mathbf{g}_2, \ldots, \mathbf{g}_M]^\mathsf{T}$, the overall estimation error of the MIMO system is also positively correlated with the number of antennas $M$.

\subsubsection{ML Estimation Using the Newton-Raphson Method}
{Since a closed-form  solution to maximizing the LLFs ${L_{\rm{p}}}( {{\bf{\bar Z}};{\bm{\eta }}} )$ and ${L_{\rm{t}}}( {{\bf{\bar Z}};{\bm{\eta }}})$ in \eqref{eq:likli} and \eqref{eq:llft} does not exist, the iterative Newton-Raphson method is next applied to obtain the ML channel estimate. However, both ${L_{\rm{p}}}( {{\bf{\bar Z}};{\bm{\eta }}} )$ and ${L_{\rm{t}}}( {{\bf{\bar Z}};{\bm{\eta }}})$  are not concave with respect to ${\bm{\eta }}$. To this end, we reparameterize ${\bm{\eta }}$ into an auxiliary vector ${{\bm{\eta }}_{{\rm{new}}}} = {[{{\bm{\varsigma }}^{\mathsf T}},\xi ]^{\mathsf T}} = {[{{{\bf{\bar g}}}^{\mathsf T}}{\rm{/}}\sigma ,{\rm{1/}}\sigma ]^{\mathsf T}}$ to make ${L_{\rm{p}}}( {{\bf{\bar Z}};{\bm{\eta }}_{{\rm{new}}}} )$ and ${L_{\rm{t}}}( {{\bf{\bar Z}};{\bm{\eta }}_{{\rm{new}}}} )$ concave with respect to ${{\bm{\eta }}_{{\rm{new}}}}$ \cite{ribeiro2006bandwidth,liu2020angular}, then the optimal ${{\bm{\hat \eta }}_{{\rm{new}}}^{}}$ can be obtained by using the following Newton-Raphson method}
\begin{eqnarray}
{{\bm \hat{\bm\eta }}}_{{\rm{new}}}^{(i)} = {{\bm \hat { \bm \eta }}}_{{\rm{new}}}^{(i-1)} + {\left({{\bf{H}}_{\rm{e}}}\right)^{ - 1}}{\nabla _{{{\bm{\eta }}_{{\rm{new}}}}}}{{{L}}_{{\rm{p/t}}}}( {{\bf{\bar Z}};{{\bm \hat{\bm\eta }}}_{{\rm{new}}}^{(i-1)}} ),
\label{eq:N-Rmethod}
\end{eqnarray}
where ${{\bf{H}}_{\rm{e}}}$ is the Hessian matrix of ${{{L}}_{{\rm{p}}}}( {{\bf{\bar Z}};{\bm{\hat \eta }}_{{\rm{new}}}^{(i-1)}} )$ or ${{{L}}_{{\rm{t}}}}( {{\bf{\bar Z}};{\bm{\hat \eta }}_{{\rm{new}}}^{(i-1)}} )$ at the point ${\bm{\hat \eta }}_{{\rm{new}}}^{(i-1)}$. After convergence, the iteration stops with the global optimum solution ${{\bm{\hat \eta }}_{{\rm{new}}}^{}}$.
On the other hand, the parameters ${[{\bf{A}}]_{i,j}} = \frac{{\sqrt 2 \left( {{{[{\bf{\bar X\bar g}}]}_i} - {\tau _j}} \right)}}{\sigma } = \sqrt 2 \left( {{{[{\bf{\bar X }}{\bm \varsigma}]}_i} - {\tau _j}\xi } \right)$, where $i = 1,2, \ldots, 2{T_{\rm{p}}}$, and $j = 1,2$, in the LLFs are associated with ${\bm{\eta }} = {[{{{\bf{\bar g}}}^{\mathsf T}},\sigma ]^{\mathsf T}}$ and ${{\bm{\eta }}_{{\rm{new}}}} = {[{{\bm{\varsigma }}^{\mathsf T}},\xi ]^{\mathsf T}} = {[{{{\bf{\bar g}}}^{\mathsf T}}{\rm{/}}\sigma ,{\rm{1/}}\sigma ]^{\mathsf T}}$. By considering the fact that a two-threshold ADC  has at least one non-zero threshold $\tau_j$, given the global optimum solution ${{\bm{\hat \eta }}_{{\rm{new}}}^{}}$, there exists a unique ${\bm{\hat \eta }} = {[{{{\bf\hat{ \bar g}}}^{\mathsf T}},\hat \sigma ]^{\mathsf T}} = [{{{\bm{\hat \varsigma }}}^{\mathsf T}}/\hat \xi ,1/\hat \xi ]$ to make ${[{\bf{\hat A}}]_{i,j}} = \frac{{\sqrt 2 \left( {{{[{\bf{\bar X}\bf\hat{\bar g}}]}_i} - {\tau _j}} \right)}}{{\hat \sigma }} = \sqrt 2 \left( {{{[{\bm{\bar X\hat \varsigma }}]}_i} - {\tau _j}\hat \xi } \right)$ valid, for $i = 1,2, \ldots ,2{T_{\rm{p}}}$, and $j = 1,2$,  so that the LLFs ${{{L}}_{{\rm{p}}}}( {{\bf{\bar Z}};{\bm{\hat \eta }}_{{\rm{}}}} )$ and ${{{L}}_{{\rm{t}}}}( {{\bf{\bar Z}};{\bm{\hat \eta }}} )$ are globally maximized, and the ML channel estimate can be obtained as ${\bf \hat{\bar g}} = {{\bm{\hat \varsigma }}}/\hat \xi$ \cite{ribeiro2006bandwidth,liu2020angular}.

\subsection{SISO Systems with Two-threshold ADCs}
\label{sec:siso}

In this subsection, the CRLB analysis is provided to illustrate how the ADC thresholds affect the error performance of the SISO system. As a special case of the model in \eqref{eq:sys_mod}, the received signal reduces to ${\bf{y}} =  g{\bf{x}} + {\bf{n}} \in {{\mathbb{C}}^{{T_{\rm{p}}} \times 1}}$, where $g$, ${\bf{x}}$ and ${\bf{n}}$ are the scalar SISO channel coefficient, the pilot vector and the noise vector, respectively. Since the rank of channel $g$ is one, all the elements in the pilot vector ${\bf{x}}$ are set to be a constant $\sqrt {{P_{\rm{s}}}} +\jmath\sqrt {{P_{\rm{s}}}} $, where $\jmath = \sqrt{-1}$. Accordingly, the quantized output in the bivariate real form ${\bf{\bar Z}}$ becomes ${\bf{\bar Z}} =  {\bf{\bar X\bar g}} + {\bf{\bar n}} \in {{\mathbb{R}}^{2{T_{\rm{p}}} \times 2}}$. By defining an intermediate parameter ${{\bm{\theta }}_i} = [\sqrt{2}({[{\bf{s}}]_i} - {\tau _1})/\sigma ,\sqrt{2}({[{\bf{s}}]_i} - {\tau _2})/\sigma ]$, where $i=1,2$, and ${\bf{s}} = \sqrt {{P_{\rm{s}}}} {[{[{\bf{\bar g}}]_1} - {[{\bf{\bar g}}]_2},{[{\bf{\bar g}}]_1} + {[{\bf{\bar g}}]_2}]^{\mathsf T}}$, the LLFs associated with the parameter set ${\bm{\theta }}_i$ are of a similar form as those in \eqref{eq:likli} and \eqref{eq:llft}.

\textit{Theorem 4:} In SISO systems with PO-ADCs and T-ADCs, the deterministic CRLBs of ${[{\bf{\bar g}}]_i}$ are respectively given by
\begin{eqnarray}
{\rm{CRL}}{{\rm{B}}_{\rm{p}}}( {{{[{\bf{\bar g}}]}_i}} )
\!\!\!\!\!&=&\!\!\!\!\!\sum\limits_{j = 1}^2 {\frac{{{\sigma ^2}}}{{8{P_{\rm{s}}}{T_{\rm{p}}}{{\left( {{{[{{\bm{\theta }}_j}]}_1} - {{[{{\bm{\theta }}_j}]}_2}} \right)}^2}}}\left([{{\bm{\theta }}_j}]_2^2c( {{{[{{\bm{\theta }}_j}]}_1}} )\right.} + \left. [{{\bm{\theta }}_j}]_1^2c( {{{[{{\bm{\theta }}_j}]}_2}} )\right),i=1,2,
\label{eq:crbsp}
\end{eqnarray}
\begin{eqnarray}
{\rm{CRL}}{{\rm{B}}_{\rm{t}}}( {{{[{\bf{\bar g}}]}_i}} ) \!\!\!\!\!&=&\!\!\!\!\! \sum\limits_{j = 1}^2 {\frac{{{\sigma ^2}}}{{8{P_{\rm{s}}}{T_{\rm{p}}}{{\left( {{{[{{\bm{\theta }}_j}]}_1} - {{[{{\bm{\theta }}_j}]}_2}} \right)}^2}}}\bigg([{{\bm{\theta }}_j}]_2^2c( {{{[{{\bm{\theta }}_j}]}_1}} )} \nonumber \\
\!\!\!\!\!&+&\!\!\!\!\! [{{\bm{\theta }}_j}]_1^2c( {{{[{{\bm{\theta }}_j}]}_2}}) \!-\! \frac{{Q\!\left( {{{[{{\bm{\theta }}_j}]}_1}} \right)Q\!\left( { - {{[{{\bm{\theta }}_j}]}_2}} \right){{[{{\bm{\theta }}_j}]}_1}{{[{{\bm{\theta }}_j}]}_2}}}{{q( {{{[{{\bm{\theta }}_j}]}_1}} )q( { - {{[{{\bm{\theta }}_j}]}_2}} )}}\bigg),i\!=\!\!1,2,
\label{eq:crbst}
\end{eqnarray}
where $c(x) = Q(x)Q( - x)/{q^2}(x)$. 

The MSE of the ML channel estimate ${ { {\bf \hat{\bar g}}}}$ is $\rm{mse_{PA,p/t}}( {{\bf\hat{ \bar g}}} )\mathop =\limits^{{\rm{a}}{\rm{.e}}{\rm{.}}} \sum\nolimits_{i = 1}^2 {{\rm{CRL}}{{\rm{B}}_{\rm{p/t}}}( {{{[{\bf{\bar g}}]}_i}} )}$.

{\textit{Proof:} See Appendix C. \hfill $\blacksquare$\\}

As shown in \eqref{eq:crbsp} and \eqref{eq:crbst}, the deterministic CRLBs  are both functions of ${{\bm{\theta }}_i},i=1,2$. The only difference lies in the last term within the parentheses $(\cdot)$, which makes ${\rm{CRL}}{{\rm{B}}_{\rm{p}}}( {{{[{\bf{\bar g}}]}_i}} ) \gtrless{\rm{  CRL}}{{\rm{B}}_{\rm{t}}}( {{{[{\bf{\bar g}}]}_i}} )$ when the product ${[{{\bm{\theta }}_j}]_1}{[{{\bm{\theta }}_j}]_2} \lessgtr 0,\forall j$. When ${[{{\bm{\theta }}_j}]_1}{[{{\bm{\theta }}_j}]_2}$ approaches to zero, the MSE of the ML channel estimate in the PO-ADC  becomes almost identical to that in the T-ADC. This occurs when the analog input ${{\rm{[}}{\bf{s}}{\rm{]}}_1}$ or ${{\rm{[}}{\bf{s}}{\rm{]}}_2}$ is much close to the threshold ${\tau _1}$ or ${\tau _2}$.

\subsection{Hybrid CRLB}
\label{sec:hcrlb}

When the channel is random and the deterministic noise variance is unknown to the BS. The MSE of the channel estimate $ {\bf \hat{\bar g}}$ is given by
\begin{eqnarray}
	{\rm{mse}}({\bf\hat{\bar g}})\mathop = {{\mathbb{E}}_{{\bf\hat{\bar g}},{\bf{\bar g}}}}\left[ {\left\| {{\bf\hat{\bar g}} - {\bf{\bar g}}} \right\|_2^2} \right].
\end{eqnarray}

It is lower bounded by the associated hybrid CRLB (HCRLB), i.e.,
\begin{eqnarray}
	{\rm{mse}}({\bf\hat{\bar g}}) \ge {\rm{tr}}\left( {{{\left[ {{{\left( {{{\bf{H}}_{{\rm{p/t}}}}({\bm{\eta }})} \right)}^{ - 1}}} \right]}_{1:2K,1:2K}}} \right),
\end{eqnarray}
where ${{{\bf{H}}_{{\rm{p/t}}}}({\bm{\eta }})}$ denotes the hybrid information matrix (HIM) in the MIMO system with PO-ADCs/T-ADCs, given by
\begin{eqnarray}
{{\bf{H}}_{{\rm{p/t}}}}({\bm{\eta }})
	\!\!\!\!\!&=&\!\!\!\!\! {{\mathbb{E}}_{{\bf{\bar Z}}|{\bm{\eta }}}}\left[ {\frac{{\partial {\rm{ln}}\left( {{\rm{Pr}}({\bf{\bar Z}},{\bm{\eta }})} \right)}}{{\partial {\bm{\eta }}}}{{\left( {\frac{{\partial {\rm{ln}}\left( {{\rm{Pr}}({\bf{\bar Z}},{\bm{\eta }})} \right)}}{{\partial {\bm{\eta }}}}} \right)}^{\mathsf T}}} \right] \nonumber \\
	\!\!\!\!\!&=&\!\!\!\!\! {{\mathbb{E}}_{{\bf{\bar Z}},{\bf{\bar g}}|\sigma }}\left[ {\frac{{\partial {\rm{ln}}\left( {{\rm{Pr(}}\left. {{\bf{\bar Z}}} \right|{\bf{\bar g}},\sigma {\rm{)}}} \right)}}{{\partial {\bm{\eta }}}}{{\left( {\frac{{\partial {\rm{ln}}\left( {{\rm{Pr(}}\left. {{\bf{\bar Z}}} \right|{\bf{\bar g}},\sigma {\rm{)}}} \right)}}{{\partial {\bm{\eta }}}}} \right)}^{\mathsf T}}} \right]\nonumber \\
	\!\!\!\!\!&+&\!\!\!\!\! {{\mathbb{E}}_{{\bf{\bar g}}|\sigma }}\left[ {\frac{{\partial {\rm{ln}}\left( {{\rm{Pr(}}{\bf{\bar g}}|\sigma {\rm{)}}} \right)}}{{\partial {\bm{\eta }}}}{{\left( {\frac{{\partial {\rm{ln}}\left( {{\rm{Pr(}}{\bf{\bar g}}|\sigma {\rm{)}}} \right)}}{{\partial {\bm{\eta }}}}} \right)}^{\mathsf T}}} \right]\nonumber \\
	\!\!\!\!\!&=&\!\!\!\!\! \underbrace{{{\mathbb{E}}_{{\bf{\bar g}}|\sigma }}\!\!\left[ {{{\bf{F}}_{\rm{p/t}}}({\bm{\eta }})} \right]}_{{{\bf{H}}_{{\rm{D,p/t}}}}({\bm{\eta }})} \!+\! \underbrace{{{\mathbb{E}}_{{\bf{\bar g}}}}\!\left[ {\frac{{\partial {\rm{ln}}\left( {{\rm{Pr(}}{\bf{\bar g}}{\rm{)}}} \right)}}{{\partial {\bm{\eta }}}}{{\left(\! {\frac{{\partial {\rm{ln}}\left( {{\rm{Pr(}}{\bf{\bar g}}{\rm{)}}} \!\right)}}{{\partial {\bm{\eta }}}}} \right)}^{\mathsf T}}} \right]}_{{{\bf{H}}_{{\rm{P,p/t}}}}({\bm{\eta }})},~~
	\label{eq:HCRLB}
\end{eqnarray}
in which, the corresponding FIM ${{{\bf{F}}_{{\rm{p/t}}}}({\bm{\eta }})}$ is given in \eqref{eq:fiser1} and \eqref{eq:fimt}, respectively. Especially, when  ${{\bf\bar{g}}}$ is white Gaussian distributed, i.e., ${{\bf\bar{g}}} \sim {\mathcal N} ({\bf{0}},\sigma^2_{\rm{g}}{\bf{I}})$, we have
\begin{eqnarray}
{{\bf{H}}_{{\rm{P,p/t}}}}({\bm{\eta }}) \!\!\!\!\!&=&\!\!\!\!\! {{\mathbb{E}}_{{\bf{\bar g}}}}\left[ {\frac{{\partial {\rm{ln}}\left( {{\rm{Pr(}}{\bf{\bar g}}{\rm{)}}} \right)}}{{\partial {\bm{\eta }}}}{{\left( {\frac{{\partial {\rm{ln}}\left( {{\rm{Pr(}}{\bf{\bar g}}{\rm{)}}} \right)}}{{\partial {\bm{\eta }}}}} \right)}^{\mathsf T}}} \right] = \left[ {\begin{array}{*{20}{c}}
	{1/\sigma _{\rm{g}}^2{\bf{I}}}&{\bf{0}}\\
	{\bf{0}}&{\bf{0}}
	\end{array}} \right],
\label{eq:Hp}
\end{eqnarray}
and
\begin{eqnarray}
{{\bf{H}}_{{\rm{D,p/t}}}}({\bm{\eta }}) = \int {{{\bf{F}}_{{\rm{p/t}}}}({\bm{\eta }})\Pr ({\bf{\bar g}})} d{\bf{\bar g}}.
\label{eq:BHIM}
\end{eqnarray}

Although the intractable multivariate integral in \eqref{eq:BHIM} prohibits an analytical evaluation on ${{\bf{H}}_{{\rm{D,p/t}}}}({\bm{\eta }})$, a numerical computation can be obtained via Monte-Carlo simulations.

\section{Joint Pilot-and-Data Channel Estimation}
\label{sec:jpd}

The CRLB analysis in Section \ref{sec:PA} indicates that a robust channel recovery from highly distorted and nonlinear quantized signals involves a large number of pilots $T_{\rm p}$. {To alleviate the pilot overhead for a more efficient data transmission, a non-pilot-aided (NPA)  scheme is considered in \cite{mezghani2018blind,das2012snr}, in which only the received data matrix is utilized for channel estimation. Inspired by this idea, we next investigate a joint pilot-and-data (JPD) estimation framework by jointly processing the received information associated with both pilot and data matrices \cite{wen2015bayes,nayebi2017semi,ding2018bayesian}.} In this way, the transmitted matrix in \eqref{eq:sys_mod} becomes ${\bf{X}} = [{{\bf{X}}_{\rm{p}}},{{\bf{X}}_{\rm{d}}}] \in {\mathbb{C}^{K \times (T_{\rm{p}}+T_{\rm{d}})}}$, which contains both the pilot matrix ${{\bf{X}}_{\rm{p}}}\in {\mathbb{C}^{K \times T_{\rm{p}}}}$ and the data matrix ${{\bf{X}}_{\rm{d}}}\in {\mathbb{C}^{K \times T_{\rm{d}}}}$, and  a new parameter set of interest ${\bm{\chi }} = {[{\bf{\bar g}}_1^{\mathsf T},{\bf{\bar g}}_2^{\mathsf T}, \cdots ,{\bf{\bar g}}_M^{\mathsf T}, \sigma]^{\mathsf T}}  \in {\mathbb{R}^{(2MK + 1) \times 1}}$ is constructed to satisfy the rank constraint in data detection phase. While JPD has been widely studied, our paper specifically provides a quantitative analysis of JPD performance in low-resolution systems, particularly focusing on PO-ADCs and T-ADCs. We further offer theoretical insights by summarizing key theorems related to these systems.

\subsection{Performance of JPD Estimation}

Suppose the symbols in ${{\bf{X}}_{\rm{d}}}$ are drawn from a given constellation ${\mathcal S}$, e.g., QPSK (${|\mathcal S|}=4$) or 16-QAM (${|\mathcal S|}=16$), then each column of ${{\bf{X}}_{\rm{d}}}$ has $L = {|\mathcal S|}^K$ possible distinct vectors ${{\bf{x}}_{l}},l=1,2,\cdots,L,$ and $\Pr ({[{{\bf{X}}_{\rm{d}}}]_{:,i}} = {{\bf{x}}_l}) = 1/L$, for $\forall i,l$. By denoting ${{{\bf{\bar Z}}}_{\rm{d}}} = [{{{\bf{\bar Z}}}_{{\rm{d,1}}}},{{{\bf{\bar Z}}}_{{\rm{d,2}}}}, \cdots ,{{{\bf{\bar Z}}}_{{\rm{d,}}M}}] \in {{\mathbb{R}}^{2{T_{\rm{d}}} \times 2M}}$, where ${{{\bf{\bar Z}}}_{{\rm{d,}}m}} \in {{\mathbb{R}}^{2{T_{\rm{d}}} \times 2}}$ is the quantized output corresponding to the analog input ${{{\bf{\bar X}}}_{\rm{d}}}{{{\bf{\bar g}}}_m}$ in the bivariate real form, for the NPA channel estimation in MIMO systems with PO-ADCs, the LLF is given by
\begin{eqnarray}
	\!\!\!\!\!&{L_{\rm{p}}}&\!\!\!\!\!( {{{{\bf{\bar Z}}}_{\rm{d}}},{\bm{\chi }}} ) = \ln \Pr ( {{{{\bf{\bar Z}}}_{\rm{d}}},{\bm{\chi }}} ) = \ln \prod\limits_{i = 1}^{{T_{\rm{d}}}} {\sum\limits_{l = 1}^L {\Pr ( { {{{{\bf{\bar z}}}_i},{\bm{\chi }}} |{{[{{\bf{X}}_{\rm{d}}}]}_{:,i}}} )\Pr ( {{{[{{\bf{X}}_{\rm{d}}}]}_{:,i}} = {{\bf{x}}_l}} )} }  \nonumber\\
	\!\!\!\!\!&=&\!\!\!\!\! \sum\limits_{i = 1}^{{T_{\rm{d}}}} {\ln \Big(\frac{1}{L}\sum\limits_{l = 1}^L {\prod\limits_{m = 1}^M {\prod\limits_{j = 1}^2 {Q\!\left( {{ \left(2[{{{\bf{\bar Z}}}_{{\rm{d,}}m}}] _{i,j}- 1\right)}{{{a_{{\rm{R,}}m,l,j}}}}} \right)} } } } Q\!\left( {{{\left(2[{{{\bf{\bar Z}}}_{{\rm{d,}}m}}]_{i + {T_{\rm{d}}},j}-1\right)}}{{{a_{{\rm{I,}}m,l,j}}}}} \right)\Big),
	\label{eq:npa}
\end{eqnarray}
where
${{{\bf{\bar z}}}_i} = {[{[{{{\bf{\bar Z}}}_{\rm{d}}}]_{i,:}},{[{{{\bf{\bar Z}}}_{\rm{d}}}]_{i + {T_{\rm{d}}},:}}]^{\mathsf T}},$
and
\[{a_{{\rm{R,}}m,l,j}} \!=\! \frac{{\sqrt 2 \left( { {\rm{Re}}({{{{{\bf{ g}}}_m^{\mathsf T}}{{{\bf{x}}_l}}}}) \!-\! {\tau _j}} \right)}}{\sigma },{a_{{\rm{I,}}m,l,j}} \!=\! \frac{{\sqrt 2 \left( { {\rm{Im}}({{{{{\bf{ g}}}_m^{\mathsf T}}{{{\bf{x}}_l}}}}) \!-\! {\tau _j}} \right)}}{\sigma }.\]

In MIMO systems with T-ADCs, the LLF  ${L_{\rm{t}}}( {{{{\bf{\bar Z}}}_{\rm{d}}},{\bm{\chi }}} )$ can be also explicitly described based on \eqref{eq:llft} and \eqref{eq:npa} as
\begin{eqnarray}
	{L_{\rm{t}}}( {{{{\bf{\bar Z}}}_{\rm{d}}},{\bm{\chi }}} ) \!\!\!\!\!&=&\!\!\!\!\! \sum\limits_{i = 1}^{{T_{\rm{d}}}} {\ln \Big(\frac{1}{L}\sum\limits_{l = 1}^L {\prod\limits_{m = 1}^M \!{\left( {1 - {{[{{{\bf{\bar Z}}}_{{\rm{d,}}m}}]}_{i,1}}} \right)Q\left( {a_{{\rm{R,}}m,l,1}} \right)} } } \nonumber\\
	\!\!\!\!\!&\cdot&\!\!\!\!\! \left( {{{[{{{\bf{\bar Z}}}_{{\rm{d,}}m}}]}_{i,1}} - {{[{{{\bf{\bar Z}}}_{{\rm{d,}}m}}]}_{i,2}}} \right)\!\left( {Q\!\left( {a_{{\rm{R,}}m,l,1}} \right) - Q\!\left( {a_{{\rm{R,}}m,l,2}} \right)} \right)\nonumber\\
	\!\!\!\!\!&\cdot&\!\!\!\!\! \left( {{{[{{{\bf{\bar Z}}}_{{\rm{d,}}m}}]}_{i,1}} - {{[{{{\bf{\bar Z}}}_{{\rm{d,}}m}}]}_{i,2}}} \right)\!\left( {Q\!\left( {a_{{\rm{R,}}m,l,1}} \right) - Q\!\left( {a_{{\rm{R,}}m,l,2}} \right)} \right) \!\!\Big).
\end{eqnarray}

According to \eqref{eq:FIM}, the FIM of the NPA estimation requires the calculation of \[{{\mathbb{E}}_{{\bf{\bar Z}}|{\bm{\chi }}}}[ {{\nabla _{\bm{\chi }}}{{L_{\rm p/t}}}( {{\bf{\bar Z}};{\bm{\chi}}} ){{( {{\nabla _{\bm{\chi }}}{{L_{\rm p/t}}}( {{\bf{\bar Z}};{\bm{\chi }}} )} )}^{\mathsf T}}} ],\] which is not analytically expressible. However, the following theorem demonstrate that under some mild conditions this FIM is equal to that of the PA estimation.

\textit{Theorem 5:} Suppose that ${{\bf{X}}_{\rm{p}}} = {{\bf{X}}_{\rm{d}}}$ and ${3^{M}} \gg {|\mathcal{S}|}^K = L $,  {given a high SNR,} we have
\begin{eqnarray}
	{{\bf{F}}_{{\rm{NPA}}}}({\bm{\chi }}) = {\bf{F}}{{\rm{}}_{{\rm{PA}}}}({\bm{\chi }}),
\end{eqnarray}
where ${{\bf{F}}_{{\rm{NPA}}}}({\bm{\chi }})$ and ${\bf{F}}{{\rm{}}_{{\rm{PA}}}}({\bm{\chi }})$ represent the FIM of NPA and PA schemes, respectively.

\textit{Proof:} See Appendix D.   \hfill $\blacksquare$\\

\textit{Theorem 5} reveals the feasibility of the NPA scheme in quantized systems. Especially, the condition ${3^{M}} \gg {|\mathcal{S}|}^K$ is usually satisfied in massive MIMO systems, because $M \gg K$. {However, the NPA based estimators are in general sensitive to the initialization \cite{das2012snr}. While, within the JPD estimation scheme, a reasonable initialization can be generated from the pilot ${{\bf{X}}_{\rm{p}}}$, and then, both the pilot ${{\bf{X}}_{\rm{p}}}$ and the detected data ${{\bf{X}}_{\rm{d}}}$ are employed for a robust channel estimate.} Since ${{\bf{X}}_{\rm{p}}}$ and ${{\bf{X}}_{\rm{d}}}$ are independent, the FIM of the JPD estimation is given by
\begin{eqnarray}
	{{\bf{F}}_{{\rm{JPD}}}}({\bm{\chi }}) = {{\bf{F}}_{{\rm{PA}}}}({\bm{\chi }}) + {{\bf{F}}_{{\rm{NPA}}}}({\bm{\chi }}),
	\label{eq:FIMJPD}
\end{eqnarray}
and the deterministic CRLB is $({{\bf{F}}_{{\rm{JPD}}}}({\bm{\chi }}))^{-1}$, in which, the associated HIM needs to be numerically calculated, similar to the one in \eqref{eq:HCRLB}.

\subsection{JPD Estimation via Expectation Maximization}
\label{sec:JPD}

In principle, the JPD estimation should operate in a turbo fashion so as to iteratively refine the estimated channel coefficients and the detected data. In this subsection, we propose EM and variational inference EM (VIEM) algorithms, dedicated for deterministic and random channels, respectively.

\subsubsection{Deterministic Channel}
The EM algorithm is considered to obtain the ML channel estimate. At the $i$-th iteration, the E-step calculates the expectation such that
\begin{eqnarray}
{\mathcal{M}} ( {{\bm{\chi }},{{{\bm{\hat \chi }}}^{(i - 1)}}} )
\!\!\!\!\!&=&\!\!\!\!\!  \sum\limits_{i = 1}^{{T_{\rm{d}}}} {{{\mathbb{E}}_{{{[{{\bf{X}}_{\rm{d}}}]}_{:,i}}|{{{\bf{\bar Z}}}_{\rm{d}}},\;{{{\bm{\hat \chi }}}^{(i - 1)}}}}\left[ {\ln \Pr \left( {{{{\bf{\bar z}}}_i},{\bm{\chi }}|{{[{{\bf{X}}_{\rm{d}}}]}_{:,i}}} \right)} \right]} +{\rm{lnPr}}({{{\bf{\bar Z}}}_{\rm{p}}},{\bm{\chi }})  \nonumber \\
\!\!\!\!\!&=&\!\!\!\!\! \sum\limits_{i = 1}^{{T_{\rm{d}}}} {\sum\limits_{l = 1}^L {\Pr \! \left(\! {{{[{{\bf{X}}_{\rm{d}}}]}_{:,i}} \!=\! {{\bf{x}}_l}|{{{\bf{\bar z}}}_i},\!\!\;{{{\bm{\hat \chi }}}^{(i - 1)}}}\! \right)\!\ln \Pr \!\left(\! {{{{\bf{\bar z}}}_i},{\bm{\chi }}|{{[{{\bf{X}}_{\rm{d}}}]}_{:,i}}} \!\right)}} + {\rm{lnPr}}({{{\bf{\bar Z}}}_{\rm{p}}},{\bm{\chi }}),
\label{eq:expectation}
\end{eqnarray}
where ${{{\bf{\bar Z}}}_{\rm{p}}}$ is the quantized output corresponding to the pilot matrix ${\bf{X}}_{\rm p}$,  ${\Pr ( {{{[{{\bf{X}}_{\rm{d}}}]}_{:,i}} = {{\bf{x}}_l}|{{{\bf{\bar z}}}_i},\;{{{\bm{\hat \chi }}}^{(i - 1)}}} )}$ is the posterior probability of ${{{[{{\bf{X}}_{\rm{d}}}]}_{:,i}} = {\bf{x}}_{l}}$, given the quantized output ${{{\bf{\bar z}}}_i}$ and the $(i-1)$-th estimate ${{{\bm{\hat \chi }}}^{(i - 1)}}$. Then the M-step maximizes ${\mathcal{M}(}{\bm{\chi }},{{{\bm{\hat \chi }}}^{(i-1)}}{\rm{)}}$ with respect to ${\bm{\chi }}$, i.e.,
\begin{eqnarray}
{{{\bm{\hat \chi }}}^{(i )}} = \arg \mathop {\max }\limits_{\bm{\chi }}  {{\mathcal{M}}}{\rm{(}}{\bm{\chi }},{{{\bm{\hat \chi }}}^{(i-1)}}{\rm{)}}.
\label{eq:maxi}
\end{eqnarray}

Invoking \eqref{eq:N-Rmethod}, we reparameterize ${\bm{\chi }}$ as ${{\bm{\chi }}_{{\rm{new}}}} = {[{\bf{\bar g}}_1^{\mathsf T}{\rm{/}}\sigma, {\bf{\bar g}}_2^{\mathsf T}{\rm{/}}\sigma , \cdots ,{\bf{\bar g}}_M^{\mathsf T}{\rm{/}}\sigma ,1/\sigma ]^{\mathsf T}}$ to make the above maximization concave with respect to ${{\bm{\chi }}_{{\rm{new}}}}$, so that, the iterative Newton-Raphson method can be employed for the global solution to \eqref{eq:maxi}. It has been proven in \cite{wu1983convergence} that  ${{{{\bm{\hat \chi }}}^{(i)}}}$ converges to a stationary point of the LLF in a non-decreasing manner. Furthermore, when the LLF is unimodal and differentiable, ${{{{\bm{\hat \chi }}}^{(i)}}}$ converges to the unique ML estimate \cite{wautelet2007comparison}.
Using the updated posterior probability ${\Pr ( {{{[{{\bf{X}}_{\rm{d}}}]}_{:,i}} = {{\bf{x}}_l}|{{{\bf{\bar z}}}_i},\;{{{\bm{\hat \chi }}}^{(i - 1)}}} )}$, the data matrix ${{\bf{X}}_{\rm{d}}}$ can be detected by finding
\begin{eqnarray}
{[{{{\bf{\hat X}}}_{\rm{d}}}]_{:,i}} = \mathop {\arg }\limits_{{{\bf{x}}_l}} \max \Pr \left( {{{[{{\bf{X}}_{\rm{d}}}]}_{:,i}} = {{\bf{x}}_l}|{{{\bf{\bar z}}}_i},\;{\bm{\hat \chi }}} \right).
\label{JPD_detect_Xd}
\end{eqnarray}

In above EM iterations, most of the posterior probabilities ${\Pr ( {{{[{{\bf{X}}_{\rm{d}}}]}_{:,i}} = {{\bf{x}}_l}|{{{\bf{\bar z}}}_i},\;{{{\bm{\hat \chi }}}^{(i - 1)}}} )},l=1,2,\ldots,L$, are sufficiently close to zero, which can be ignored when calculating the weighted LLF in \eqref{eq:expectation}. This pruning step effectively speeds up the algorithm, especially when $L = {|\mathcal S|}^K$ is prohibitive.
Moreover, the complexity of calculating the gradient and the Hessian matrix in \eqref{eq:maxi} grows exponentially with  $M \times K$. To address this issue, a parallel EM (PEM) algorithm is further proposed, where the EM steps in \eqref{eq:expectation} and  \eqref{eq:maxi} are decomposed into $m$ subproblems, i.e.,
\begin{eqnarray}
{\mathcal{M}} ( {{{\bm{\eta }}_m},{{{\bm{\hat \chi }}}^{(i - 1)}}} )
\!\!\!\!&=&\!\!\!\!\!  \sum\limits_{i = 1}^{{T_{\rm{d}}}} {{{\mathbb{E}}_{{{[{{\bf{X}}_{\rm{d}}}]}_{:,i}}|{{{\bf{\bar Z}}}_{\rm{d}}},\;{{{\bm{\hat \chi }}}^{(i - 1)}}}}\left[ {\ln \Pr \left( {{{{\bf{\bar z}}}_i},{{\bm{\eta }}_m}|{{[{{\bf{X}}_{\rm{d}}}]}_{:,i}}} \right)} \!\right]} + {\rm{lnPr}}({{{\bf{\bar Z}}}_{\rm{p}}},{{\bm{\eta }}_m}),
\label{eq:PEM1}\\
{\bm{\hat \eta }}_m^{(i )}  \!\!\!\!&=&\!\!\!\! \arg \mathop {\max }\limits_{{\bm{\eta }}_m}  {{\mathcal{M}}}{\rm{(}}{{\bm{\eta }}_m},{{{\bm{\hat \chi }}}^{(i-1)}}{\rm{)}},
\label{eq:PEMs}
\end{eqnarray}
where ${{\bm{\eta }}_m} = {[{\bf{\bar g}}_m^{\mathsf T},\sigma ]^{\mathsf T}}$ and $m$ = 1, 2, $\cdots$, $M$. All the channel vectors $\{ {{{\bf{\bar g}}}_m}\} _{m = 1}^M$ are estimated by solving the above $m$ subproblems in parallel so as to reduce the runtime. However, the reduced complexity of the PEM algorithm comes at the cost of diminished estimation accuracy, as demonstrated in Appendix E. Interestingly, this accuracy degradation is confined to the quantized system; in the full-resolution system, parallel estimation of $\{ {{{\bf{\bar g}}}_m}\} _{m = 1}^M$ does not negatively impact the resulting MSE. Building upon the structure of PEM, we introduce a refinement by dividing $\{ {{{\bf{\bar g}}}_m}\} _{m = 1}^M$ into $N$ groups ($N<M$), which are then estimated concurrently using the EM algorithm. This approach, termed the grouped PEM (GPEM) algorithm, leverages more effective sampling, thus improving the Fisher information measure for noise variance. GPEM strikes a desirable balance between computational efficiency and estimation accuracy. The JPD channel estimation for a deterministic channel is summarized in Algorithm \ref{alg:JPD_deterministic}.

\begin{algorithm}
	\caption{EM algorithm for deterministic channel in the JPD scheme}
	\label{alg:JPD_deterministic}{
	\renewcommand{\algorithmicrequire}{\textbf{Input:}}
	\renewcommand{\algorithmicensure}{\textbf{Output:}}
	\begin{algorithmic}[1]
		\REQUIRE the quantized output ${{{\bf{\bar Z}}}_{\rm{p}}}$ and ${{{\bf{\bar Z}}}_{\rm{d}}}$, and the pilot matrix ${\bf{X}}_{\rm p}$;
		\STATE Initialize the estimate ${\bm{\chi }}^{(1)} = {[{\bf{\bar g}}_1^{\mathsf T},{\bf{\bar g}}_2^{\mathsf T}, \cdots ,{\bf{\bar g}}_M^{\mathsf T}, \sigma]^{\mathsf T}}$;
		\FOR {the $i$-th iteration of EM algorithm}
		\STATE Calculate the E-step expectation ${\mathcal{M}} ( {{\bm{\chi }},{{{\bm{\hat \chi }}}^{(i - 1)}}} )$ as Eq. \eqref{eq:expectation};
		\STATE Maximizes the M-step ${\mathcal{M}(}{\bm{\chi }},{{{\bm{\hat \chi }}}^{(i-1)}}{\rm{)}}$ with respect to ${\bm{\chi }}$ as Eq. \eqref{eq:maxi};
		\STATE Reparameterize ${\bm{\chi }}$ as ${\bm{\chi }}_{\rm{new}}= {[{\bf{\bar g}}_1^{\mathsf T}{\rm{/}}\sigma, {\bf{\bar g}}_2^{\mathsf T}{\rm{/}}\sigma , \cdots ,{\bf{\bar g}}_M^{\mathsf T}{\rm{/}}\sigma ,1/\sigma ]^{\mathsf T}}$ to make the above maximization concave with respect to ${\bm{\chi }}_{\rm{new}} $ by invoking \eqref{eq:N-Rmethod};
		\STATE Employ the Newton-Raphson method to obtain the global solution of ${\bm{\chi }}_{\rm{new}}$;
		\STATE Detect data matrix $\mathbf{X}_\mathrm{d}$ by Eq. \eqref{JPD_detect_Xd};
		\ENDFOR
		\ENSURE Estimated channel coefficients ${\bm{\chi }}$, data matrix $\mathbf{X}_\mathrm{d}$.
	\end{algorithmic}}
\end{algorithm}

\subsubsection{Random Channel}

It is known that the posterior mean Bayesian estimators achieve the optimal performance for random channels \cite{kay1993fundamentals,wen2015bayes}. To this end, we next propose the variational inference EM (VIEM) algorithm.

Before proceeding to the development, we use a new representation of the quantized output to facilitate the analysis. According to \eqref{eq:tern}, the quantized output of a T-ADC is
\begin{eqnarray}
	\tilde z = {{\mathcal Q}_{\rm{t}}}(u + n) = \left\{ {\begin{array}{*{20}{c}}
		{ - \alpha\delta ,}&{u + n  \in \left( { - \infty ,{\tau _1}} \right);}\\
		{0,}&{ u + n \in \left[ {{\tau _1},{\tau_2}} \right);}\\
		{\alpha\delta ,}&{u + n \in \left[ {{\tau _2},\infty } \right),}
		\end{array}} \right.
\end{eqnarray}
where ${\tau _1} =  - {\tau _2} < 0$,  $\delta > 0$ denotes the quantization label, and
\begin{eqnarray}
	\alpha  = \sqrt {\frac{{{\mathbb{E[}}{{| {u + n} |}^2}{\rm{]}}}}{{4{\delta ^2}Q\!\left( {\sqrt {\frac{{2\tau _1^2}}{{{\mathbb{E[}}{{| {u + n} |}^2}{\rm{]}}}}} } \right)}}}
\end{eqnarray}
is a scaling factor to make the variance of $\tilde z$ equal to ${\mathbb{E[}}{| {u + n} |^2}{\rm{]}}$.

While for the PO-ADC in \eqref{eq:poadc}, we have
\begin{eqnarray}
	{{\tilde z}_i} = {{\mathcal Q}_{\rm{p}}}(u + {n_i}) = \left\{ {\begin{array}{*{20}{c}}
		{ - \alpha\delta ,}&{u + {n_i} < {\tau _i}};\\
		{\alpha\delta ,}&{u + {n_i} \ge {\tau _i}}.
		\end{array}} \right.
\end{eqnarray}
In this way, the quantized output of ${{\bf{Y}}}$ in \eqref{eq:sys_mod} becomes
\begin{eqnarray}
	{\bf{\tilde Z}} = [{{{\bf{\tilde Z}}}_{\rm{p}}},{{{\bf{\tilde Z}}}_{\rm{d}}}] = {{\mathcal Q}_{{\rm{p/t}}}}\left( [{{\bf{Y}}_{\rm{p}}},{{\bf{Y}}_{\rm{d}}}] \right)= {{\mathcal Q}_{{\rm{p/t}}}}\left( {{\bf{H}}{{\bf{X}}} + {\bf{N}}} \right),
	\label{eq:sysba}
\end{eqnarray}
where ${{\mathcal Q}_{{\rm{p/t}}}}( \cdot )$ denote that the quatilization is applied component-wise on the real/imaginary part of the analog input, and ${{{\bf{\tilde Z}}}_{\rm{p}}} \in {{\mathbb C}^{M \times {T_{\rm{p}}}}}$ and  ${{{\bf{\tilde Z}}}_{\rm{d}}} \in {{\mathbb C}^{M \times {T_{\rm{d}}}}}$ are the quantized output corresponding to the pilot phase analog input ${{\bf{Y}}_{\rm{p}}}$ and data phase analog input ${{\bf{Y}}_{\rm{d}}}$, respectively. The vectorization of \eqref{eq:sysba} is given by
\begin{eqnarray}
	{\bf{\tilde z}} = {[{\bf{\tilde z}}_{\rm{p}}^{\mathsf T},{\bf{\tilde z}}_{\rm{d}}^{\mathsf T}]^{\mathsf T}} = {{\mathcal Q}_{{\rm{p/t}}}}\left( [{\bf{\tilde y}}_{\rm{p}}^{\mathsf T},{\bf{\tilde y}}_{\rm{d}}^{\mathsf T}]^{\mathsf T} \right) =  {{\mathcal Q}_{{\rm{p/t}}}} \left( {{\bf{\tilde X\tilde h}} + {\bf{\tilde n}}} \right),
\end{eqnarray}
where ${{{\bf{\tilde z}}}_{{\rm{p/d}}}} = {\rm{vec(}}{{{\bf{\tilde Z}}}_{{\rm{p/d}}}}{\rm{)}}$, ${{{\bf{\tilde y}}}_{{\rm{p/d}}}} = {\rm{vec(}}{{{\bf{\tilde Y}}}_{{\rm{p/d}}}}{\rm{)}}$, ${\bf{\tilde X}} = {{\bf{X}}^{\mathsf T}} \otimes {{\bf{I}}_M}$,  ${\bf{\tilde h}} = {\rm{vec(}}{\bf{H}}{\rm{)}}$ and ${\bf{\tilde n}} = {\rm{vec(}}{\bf{N}}{\rm{)}}$.

Suppose the channel prior as ${\bf{\tilde h}} \sim {\mathcal{  CN(}}{\bf{0}}{\rm{,}}{\bm{\Lambda }}{\rm{)}} $, where ${\bm{\Lambda }} = {\rm{diag(}}{\bm{\lambda }}{\rm{)}}$ and ${\bm{\lambda }} = [{\lambda _1},{\lambda _2}, \ldots ,{\lambda _{MK}}]^{\mathsf T}$ denotes the hyperparameter vector, and define the parameter set ${\bf{\Theta }} = \{{{\bf{X}}_{\rm{d}}}{\rm{,}}{\bm{\lambda }} \}$, the E-step of the proposed VIEM algorithm calculates

\begin{eqnarray}
	Q\left( {{\bf{\Theta }}|{{\bf{\Theta }}^{(p - 1)}}} \right) \!\!\!\!\!&=&\!\!\!\!\! {{\mathbb{E}}_{{\bf{\tilde h}}|{\bf{\tilde z}};{{\bf{\Theta }}^{(p - 1)}}}}\left[ {\ln \Pr \left( {{\bf{\tilde z}},{\bf{\tilde h}};{\bf{\Theta }}} \right)} \right] \nonumber\\
	\!\!\!\!\!&=&\!\!\!\!\! \ln \Pr \left( {{\bf{\tilde z}}|{\bf{\tilde y}}} \right) + {{\mathbb{E}}_{{\bf{\tilde h}}|{\bf{\tilde z}};{{\bf{\Theta }}^{(p - 1)}}}}\left[\ln \Pr \left( {{\bf{\tilde y}}|{\bf{\tilde h}}} \right)\right. + \left.\ln \Pr \left( {{\bf{\tilde h}}|{\bm{\lambda }}} \right)\right] + \Pr \left( {\bm{\lambda }} \right).
	\label{eq:E-step}
\end{eqnarray}

In the Bayesian inference framework, the non-informative prior of ${\bm{\lambda }}$ is considered \cite{singh2013bayes}, which means that the last term $\Pr \left( {\bm{\lambda }} \right)$ in \eqref{eq:E-step} can be dropped. Then the M-step of VIEM computes the maximum of $Q( {{\bf{\Theta }}|{{\bf{\Theta }}^{(p - 1)}}} )$ such that
\begin{eqnarray}
	\left( {{\bf{X}}_{\rm{d}}^{(p)},{{\bm{\lambda }}^{(p)}}} \right) = \mathop {\arg \max }\limits_{[{{\bf{X}}_{\rm{d}}]_{i,j}} \in {{\mathcal{S}}},{\bm{\lambda }}} Q\left( {{\bm{\Theta }}|{{\bm{\Theta }}^{(p - 1)}}} \right).
	\label{eq:VIEM-M-step}
\end{eqnarray}

At the $p$-th iteration, the E-step requires the exact posterior probability $\Pr ({\bf{\tilde h}}|{\bf{\tilde z}};{{\bf{\Theta }}^{(p - 1)}})$ to compute the channel estimate ${\bf{\tilde h}}^{(p)} = {{\mathbb{E}}_{{\bf{\tilde h}}|{\bf{\tilde z}};{{\bf{\Theta }}^{(p - 1)}}}}[{\bf{\tilde h}}]$, which unfortunately is not analytically expressible. To attain the approximate inference of $\Pr ({\bf{\tilde h}}|{\bf{\tilde z}};{{\bf{\Theta }}^{(p - 1)}})$, the variational Bayesian method \cite{tzikas2008variational} is applied by taking into account the relationship that

\begin{eqnarray}
	\ln \Pr \left( {{\bf{\tilde z}}} \right) = F( {I} ) + KL( {I||\Pr } ), \label{eq:Fq}\\
	F( I ) = \int {I( {\mathcal{P}} )\ln \frac{{\Pr ( {{\bf{\tilde z}},{\mathcal{P;}}{\bf{\Theta }}} )}}{{I( {\mathcal{P}} )}}} d{\mathcal{P}},\\
	KL( {I||\Pr } ) =  - \int {I( {\mathcal{P}} )\ln \frac{{\Pr \left( {{\mathcal{P|}}{\bf{\tilde z}}{\rm{;}}{\bf{\Theta }}} \right)}}{{I( {\mathcal{P}} )}}} ,
\end{eqnarray}
where ${\mathcal{P}} = \{ {\bf{\tilde y}},{\bf{\tilde h}}\} $, and $KL( {I||\Pr } )$ denotes the Kullback-Leibler divergence between ${I( {\mathcal{P}} )}$ and ${\Pr ( {{\mathcal{P|}}{\bf{\tilde z}}{\rm{;}}{\bf{\Theta }}} )}$. Since $KL( {I||\Pr } ) \ge 0$, we can choose a member from a certain family of distribution ${I( {\mathcal{P}} )}$ that closely approximates the true posterior ${\Pr \left( {{\mathcal{P|}}{\bf{\tilde z}}{\rm{;}}{\bf{\Theta }}} \right)}$ when $KL( {I||\Pr } ) \to 0$. In fact, from \eqref{eq:Fq}, the minimization of $KL( {I||\Pr } )$ is mathematically equivalent to the maximization of $F(I)$. Moreover, we know from the well-established mean-field theory that the inference ${I( {\mathcal{P}} )}$ can be factorized into several disjoint components, i.e., $I( {\mathcal{P}} ) = I({\bf{\tilde y}})I({\bf{\tilde h}})$. Based on the calculus of variations, the maximum of $F(I)$ is computed over the factorized inference $I( {\mathcal{P}})$, to give
\begin{eqnarray}
	I({{\mathcal{P}}_i}){\rm{ = }}\frac{{\exp \left( {\ln {{\left\langle {p({\bf{Z}},{\mathcal{P}})} \right\rangle }_{j \ne i}}} \right)}}{{\int {\exp \left( {\ln {{\left\langle {p({\bf{Z}},{\mathcal{P}})} \right\rangle }_{j \ne i}}} \right)d{{\mathcal{P}}_i}} }},
	\label{eq:varia}
\end{eqnarray}
where ${\left\langle  \cdot  \right\rangle _{j \ne i}}$ denotes the expectation over all the distributions $I({{\mathcal{P}}_j})$, for $\forall j$ except $i$.

\textit{\textbullet Calculation of $I({\bf{\tilde y}})$:} According to \eqref{eq:varia}, we have
\begin{eqnarray}
	I( {{{{\bf{\tilde y}}}}} )  \!\!\!\!&\propto&\!\!\!\!  \exp \left( {\ln \Pr\left( {{{{\bf{\tilde z}}}}|{{{\bf{\tilde y}}}}} \right) + {{\left\langle {\ln \Pr\left( {{{{\bf{\tilde y}}}}|{{{\bf{\tilde h}}}}} \right)} \right\rangle }_{I\left( {{{{\bf{\tilde h}}}}} \right)}}} \right) \nonumber\\
	\!\!\!\! &\propto& \!\!\!\! \Pr\left( {{{{\bf{\tilde z}}}}|{{{\bf{\tilde y}}}}} \right)\exp \left( {{{\left\langle {\ln \Pr\left( {{{{\bf{\tilde y}}}}|{{{\bf{\tilde h}}}}} \right)} \right\rangle }_{I\left( {{{{\bf{\tilde h}}}}} \right)}}} \right) \nonumber\\
	\!\!\!\! &\propto& \!\!\!\! {{\mathcal{I}}_{{{{\bf{\tilde y}}}}}}( {{{\mathcal{D}}_{{{{\bf{\tilde z}}}}}}} )\exp \left( {{{\left\langle { - \frac{1}{{{\sigma ^2}}}\left\| {{{{\bf{\tilde y}}}} - {{{\bf{\tilde X}}}}{{{\bf{\tilde h}}}}} \right\|_2^2} \right\rangle }_{I\left( {{{{\bf{\tilde h}}}}} \right)}}} \right) \nonumber\\
	\!\!\!\! &\propto& \!\!\!\! {{\mathcal{I}}_{{{{\bf{\tilde y}}}}}}( {{{\mathcal{D}}_{{{{\bf{\tilde z}}}}}}} )\exp \left( \!\!{ - \frac{1}{{{\sigma ^2}}}\left\| {{{{\bf{\tilde y}}}} - {{{\bf{\tilde X}}}}{{\left\langle {{{{\bf{\tilde h}}}}} \right\rangle }_{I\left( {{{{\bf{\tilde h}}}}} \right)}}} \right\|_2^2} \right),
	\label{eq:infer_y}
\end{eqnarray}
where the set ${{\mathcal{D}}_{{{{\bf{\tilde z}}}}}} =  \{ {\bf{a}} + {\bf{b}}\jmath |{{\bf{l}}_{\rm{R}}} < {\bf{a}} < {{\bf{u}}_{\rm{R}}},{{\bf{l}}_{\rm{I}}} < {\bf{b}} < {{\bf{u}}_{\rm{I}}}\} $, ${{\bf{l}}_{\rm{R/I}}}$ and ${{\bf{u}}_{\rm{R/I}}}$ correspond to the lower limit and the upper one of the quantized region of the real/imaginary part in ${{{{\bf{\tilde z}}}}}$, and ${{\mathcal{I}}_{{{{\bf{\tilde y}}}}}}( {{{\mathcal{D}}_{{{{\bf{\tilde z}}}}}}} )$ is an indicator which equals to 1 if ${{{\bf{\tilde y}}}} \in {{\mathcal{D}}_{{{{\bf{\tilde z}}}}}}$ or 0 otherwise. For example, when $u + n  \in \left( { - \infty ,{\tau _1}} \right)$, the lower and upper limits are $-\infty$ and $\tau _1$, respectively. It is recognized from \eqref{eq:infer_y} that ${[{{{\bf{\tilde y}}}}]_i}$ is subject to a truncated multivariate complex Gaussian distribution with mean ${[ {{{\bf{\tilde X}}}}{\langle {{{{\bf{\tilde h}}}}} \rangle _{I( {{{{\bf{\tilde h}}}}} )}}]_i}$ and variance $\sigma^2$. Therefore, the first moment ${\left\langle {{{[{{{\bf{\tilde y}}}}]}_i}} \right\rangle _{I({{{\bf{\tilde y}}}})}}$ can be expanded as
\begin{eqnarray}
\langle {{{[{{{\bf{\tilde y}}}}]}_i}} \rangle _{I({{{\bf{\tilde y}}}})}
\!\!\!\!\!&=&\!\!\!\!\! \frac{\sigma^{(p)} }{{\sqrt 2 }}\!\cdot\!\frac{{q ({\mu _{\rm{R}}}) - q ({\upsilon _{\rm{R}}})}}{{Q ({\upsilon _{\rm{R}}}) - Q ({\mu _{\rm{R}}})}} \!+\! {\mathop{\rm Re}\nolimits} \!\left( {{{\left[ { {{{\bf{\tilde X}}}}{{\left\langle {{{{\bf{\tilde h}}}}} \right\rangle }_{I\left( {{{{\bf{\tilde h}}}}} \right)}}} \right]}_i}} \right) \nonumber\\
\!\!\!\!\!&+&\!\!\!\!\! \jmath\left(\frac{\sigma^{(p)} }{{\sqrt 2 }}\!\cdot\!\frac{{q ({\mu _{\rm{I}}}) - q ({\upsilon _{\rm{I}}})}}  {{Q ({\upsilon _{\rm{I}}}) - Q ({\mu _{\rm{I}}})}} \!+\! {\mathop{\rm Im}\nolimits} \!\left( {{{\left[ { {{{\bf{\tilde X}}}}{{\left\langle {{{{\bf{\tilde h}}}}} \right\rangle }_{I\left( {{{{\bf{\tilde h}}}}} \right)}}} \right]}_i}} \right)\right),~~~~
\label{eq:yq}
\end{eqnarray}
where
\[{\mu _{\rm{R}}} = \sqrt 2 ({\mathop{\rm Re}\nolimits} ({[ {{{\bf{\tilde X}}}}{\langle {{{{\bf{\tilde h}}}}} \rangle _{I({{{\bf{\tilde h}}}})}}]_i}) - {[{{\bf{l}}_{\rm{R}}}]_i})/\sigma^{(p)}, \] \[{\mu _{\rm{I}}} = \sqrt 2 ({\mathop{\rm Im}\nolimits} ({[ {{{\bf{\tilde X}}}}{\langle {{{{\bf{\tilde h}}}}} \rangle _{I({{{\bf{\tilde h}}}})}}]_i}) - {[{{\bf{l}}_{\rm{I}}}]_i})/\sigma^{(p)} ,\] \[{\upsilon _{\rm{R}}} = \sqrt 2 ({\mathop{\rm Re}\nolimits} ({[ {{{\bf{\tilde X}}}}{\langle {{{{\bf{\tilde h}}}}} \rangle _{I({{{\bf{\tilde h}}}})}}]_i}) - {[{{\bf{u}}_{\rm{R}}}]_i})/\sigma^{(p)} ,\] \[{\upsilon _{\rm{I}}} = \sqrt 2 ({\mathop{\rm Im}\nolimits} ({[ {{{\bf{\tilde X}}}}{\langle {{{{\bf{\tilde h}}}}} \rangle _{I({{{\bf{\tilde h}}}})}}]_i}) - {[{{\bf{u}}_{\rm{I}}}]_i})/\sigma^{(p)}. \]

\textit{\textbullet Calculation of $I({\bf{\tilde h}})$: } Similar to  \eqref{eq:infer_y}, we have
\begin{eqnarray}
	I( {{{{\bf{\tilde h}}}}} ) \!\!\!\! &\propto& \!\!\!\! \exp \left( {{{\left\langle {\ln \Pr\left( {{{{\bf{\tilde y}}}}|{{{\bf{\tilde h}}}}} \right)} \right\rangle }_{I\left( {{{{\bf{\tilde y}}}}} \right)}} \!\!\!\!+\! {{\left\langle {\ln \Pr\left( {{{{\bf{\tilde h}}}}|{\bm{\lambda }}} \right)} \right\rangle }_{I\left( {\bm{\lambda }} \right)}}} \right) \nonumber\\
	\!\!\!\! &\propto& \!\!\!\! \exp\! \left(\!\! {{{\left\langle { - \frac{1}{{{\sigma ^2}}}\left\| {{{{\bf{\tilde y}}}} - {{{\bf{\tilde X}}}}{{{\bf{\tilde h}}}}} \right\|_2^2} \right\rangle }_{I\left( {{{{\bf{\tilde y}}}}} \right)}} \!\!\!\!\!+\! {{\left\langle { - {\bf{\tilde h}}^{\mathsf H}{{\bf{\Lambda }}^{ - 1}}{{{\bf{\tilde h}}}}} \right\rangle }_{I\left( {\bm{\lambda }} \right)}}} \right) \nonumber\\
	\!\!\!\! &\propto& \!\!\!\! \exp\! \left( { - {{\left( {{{{\bf{\tilde h}}}} - {{\bf{m}}}} \right)}^{\mathsf H}}{\bf{\Omega }}^{ - 1}\left( {{{{\bf{\tilde h}}}} - {{\bf{m}}}} \right)} \right),
	\label{eq:qh1}
\end{eqnarray}
where
\begin{eqnarray}
{{\bf{m}}} = {{{{\bf{\tilde h}}}^{(p)}}} = \frac{1}{{{(\sigma^{(p)}) ^2}}}{{\bf{\Omega }}}{\bf{\tilde X}}^{\mathsf H}{\left\langle {{{{\bf{\tilde y}}}}} \right\rangle _{I\left( {{{{\bf{\tilde y}}}}} \right)}},
\label{eq:me1}
\end{eqnarray}
and
\begin{eqnarray}
{{\bf{\Omega }}} = {\left( {\frac{1}{{({\sigma^{(p)}) ^2}}}{\bf{\tilde X}}^{\mathsf H}{{{\bf{\tilde X}}}} + \left\langle {\bm{\Lambda }} \right\rangle _{I\left( {\bm{\lambda }} \right)}^{ - 1}} \right)^{ - 1}}.
\label{eq:co1}
\end{eqnarray}

Observe from \eqref{eq:qh1} that $I({{{\bf{\tilde h}}}})$ is Gaussian distributed with mean ${{\bf{m}}}$ and covariance ${{\bf{\Omega }}}$. The calculation of $I({{{\bf{\tilde h}}}})$ requires ${\langle {{{{\bf{\tilde y}}}}} \rangle _{I({{{\bf{\tilde y}}}})}}$ in \eqref{eq:yq} and $\left\langle {\bm{\Lambda }} \right\rangle _{I\left( {\bm{\lambda }} \right)} = {\rm{diag}}\left( {{\bm{\lambda }}^{(p-1)}} \right)$. Because the above inferences have no explicit solutions, a consistent inference is attained by cycling through all these factors in turn within the EM iterations and updating each with the revised one. In the M-step, the ML estimate of the hyperparameter vector $\bm \lambda$
is given by
\begin{eqnarray}
	{{\bm{\lambda }}^{(p)}} \!\!\!\!&=&\!\!\!\! \mathop {\arg \max }\limits_{\bm{\lambda }} {{\rm{E}}_{{\bf{\tilde h}}|{\bf{\tilde z}};{{\bf{\Theta }}^{(p - 1)}}}}\left[ {\ln \Pr \left( {{\bf{\tilde h}}|{\bm{\lambda }}} \right)} \right] = \mathop {\arg \min }\limits_{\bm{\lambda }} {{\rm{E}}_{{\bf{\tilde h}}|{\bf{\tilde z}};{{\bf{\Theta }}^{(p - 1)}}}}\left[ {{{{\bf{\tilde h}}}^{\mathsf H}}{{\bf{\Lambda }}^{ - 1}}{\bf{\tilde h}}} \right] + \ln \left| {\bf{\Lambda }} \right| \nonumber\\
	\!\!\!\!&=&\!\!\!\! {\rm{diag}}\left( {\bf{K}} \right),
	\label{eq:cal_lam}
\end{eqnarray}
where ${\bf{K}} = {\bf{\Omega }} + {\bf{m}}{{\bf{m}}^{\mathsf H}}$. Meanwhile, the data matrix ${\bf{X}}_{\rm{d}}$ can be detected from the following optimization,
\begin{eqnarray}
	{\bf{X}}_{\rm{d}}^{(p)} \!\!\!\!\!&=&\!\!\!\!\! \mathop {\arg \max }\limits_{[{{\bf{X}}_{\rm{d}}]_{i,j}} \in {{\mathcal{S}}}} {{\rm{E}}_{{\bf{\tilde h}}|{\bf{\tilde z}};{{\bf{\Theta }}^{(p - 1)}}}}\left[ {\ln \Pr \left( {{\bf{\tilde y}}|{\bf{\tilde h}}} \right)} \right] = \mathop {\arg \min }\limits_{[{{\bf{X}}_{\rm{d}}]_{i,j}} \in {{\mathcal{S}}}} {{\rm{E}}_{{\bf{\tilde h}}|{\bf{\tilde z}};{{\bf{\Theta }}^{(p - 1)}}}}\left[ {\left\| { \langle{{\bf{\tilde y}}_{\rm{d}}} \rangle _{I({{{\bf{\tilde y}}}})} - {{{\bf{\tilde X}}}_{\rm{d}}}{\bf{\tilde h}}} \right\|_2^2} \right] \nonumber\\
	\!\!\!\!\!\!\!&=&\!\!\!\!\! \mathop {\arg \min }\limits_{[{{\bf{X}}_{\rm{d}}]_{i,j}} \in {{\mathcal{S}}}} \left\{ {{\rm{Tr}}\left( {{{{\bf{\tilde X}}}_{\rm{d}}}{\bf{K\tilde X}}_{\rm{d}}^{\mathsf H}} \right) + \left\| {\langle{{\bf{\tilde y}}_{\rm{d}}} \rangle _{I({{{\bf{\tilde y}}}})} - {{{\bf{\tilde X}}}_{\rm{d}}}{\bf{m}}} \right\|_2^2} \right\}\!.~~~~
	\label{eq:cal_Xd}
\end{eqnarray}

However, the minimization in \eqref{eq:cal_Xd} is usually computationally unaffordable. For example, when the BS supports $K=6$ users and the 16 QAM constellation (${|\mathcal S|}=16$)  is adopted, there exists $16^{6}$ candidates for each column in ${{\bf{X}}_{{\rm{d}}}}$. To ease its practical use, we first detect ${\bf{X}}_{\rm{d}}^{(p)}$ from \eqref{eq:cal_Xd} without accounting for the constraint ${[{{\bf{X}}_{\rm{d}}]_{i,j}} \in {{\mathcal{S}}}}$, and then, perform the hard decision on ${\bf{X}}_{\rm{d}}^{(p)}$ to yield the mapped symbols. Since the objective functions in \eqref{eq:cal_lam} and \eqref{eq:cal_Xd} are decoupled, ${{\bm{\lambda }}^{(p)}}$ and ${\bf{X}}_{\rm{d}}^{(p)}$ can be estimated separately. Given ${{{\bf{\tilde h}}}^{(p)}}$, the last step in the M-step is to estimate the unknown noise variance, i.e.,
\begin{eqnarray}
	{\sigma ^{(p)}} = \mathop {\arg \max }\limits_\sigma  {\rm{lnPr}}({\bf{\tilde z}},{{{\bf{\tilde h}}}^{(p)}},\sigma ).
	\label{eq:MLvar}
\end{eqnarray}
The objective function in \eqref{eq:MLvar} is concave with respect to $1/\sigma $, so that it can be optimized via the Newton-Raphson method at a low computation cost. The proposed VIEM algorithm iteratively updates the required statistics from \eqref{eq:infer_y} to \eqref{eq:MLvar} until convergence. Since a good initialization is key to ensure the local minimum is sufficiently approximate to the global
one \cite{winn2005variational}, the ZF estimate is used as the initial channel guess, i.e., ${{\bf{H}}^{(1)}} = {{{\bf{\tilde Z}}}_{\rm{p}}}{{\bf{X}}^{\mathsf H}}{({\bf{X}}{{\bf{X}}^{\mathsf H}})^{ - 1}}$, ${\sigma ^{(1)}}$ is randomly initialized at the system noise level and ${\rm{diag}}\left( {{\bm{\lambda }}^{(1)}} \right) = {\bf I}_{MK}$.
We summarize the JPD channel estimation for a random channel in Algorithm \ref{alg:JPD_random}.

\begin{algorithm}
	\caption{VIEM algorithm for random channel in the JPD scheme}
	\label{alg:JPD_random}{
	\renewcommand{\algorithmicrequire}{\textbf{Input:}}
	\renewcommand{\algorithmicensure}{\textbf{Output:}}
	\renewcommand{\algorithmiccomment}[1]{\hfill $\triangleright$ #1}
	\begin{algorithmic}[1]
	\REQUIRE the quantized output ${{{\bf{\tilde Z}}}_{\rm{p}}}$ and ${{{\bf{\tilde Z}}}_{\rm{d}}}$, and the pilot matrix ${\bf{X}}_{\rm p}$;
	\STATE Initialize the parameter set ${\bf{\Theta }}^{(1)} = \{{\bf{X}}^{(1)}_{\rm{d}}{\rm{,}}{\bm{\lambda }}^{(1)} \}$ and the noise variance $\sigma^{(1)}$;
	\FOR {the $p$-th iteration of VIEM algorithm}
	\STATE Obtain the approximate inference of $\Pr ({\bf{\tilde h}}|{\bf{\tilde z}};{{\bf{\Theta }}^{(p - 1)}})$ by Eq. \eqref{eq:Fq} - \eqref{eq:co1};
	\STATE Calculate the E-step expectation $Q( {{\bf{\Theta }}|{{\bf{\Theta }}^{(p - 1)}}})$ based on the obtained $\Pr ({\bf{\tilde h}}|{\bf{\tilde z}};{{\bf{\Theta }}^{(p - 1)}})$ as Eq. \eqref{eq:E-step};
	\STATE Compute the M-step maximum $Q( {{\bf{\Theta }}|{{\bf{\Theta }}^{(p - 1)}}})$ as Eq. \eqref{eq:VIEM-M-step}
	\STATE Obtain the estimated hyperparameter vector ${\bm \lambda}^{(p)}$ as Eq. \eqref{eq:cal_lam};
	\STATE Detect the data matrix $\mathbf{X}_\mathrm{d}^{(p)}$ by \eqref{eq:cal_Xd};
	\STATE Estimate the unknown noise variance $\sigma^{(p)}$ as Eq. \eqref{eq:MLvar};
	\ENDFOR
	\ENSURE Estimated channel coefficients ${\bm{\lambda}}$, data matrix $\mathbf{X}_\mathrm{d}$ and noise variance $\sigma$.
	\end{algorithmic}}
\end{algorithm}

\subsubsection{Complexity Analysis}
The computational loads of the proposed EM-based algorithms are summarized as follows.
\begin{enumerate}
	\item \textbf{For deterministic channel:}
	\begin{itemize}
		\item In each iteration of the EM algorithm, the computational complexity for the E-step is $\mathcal{O}(KM)$, while the M-step involves the computation of ${\bm{\chi}}$ and $\mathbf{X}_\mathrm{d}$, with complexities of  $\mathcal{O}(M^3K^3)$ and $\mathcal{O}(K^3T_\mathrm{d})$, respectively. Overall, the complexity of EM algorithm is dominated by the Hessian matrix inversion for calculating ${\bm{\chi}}$, which has a complexity of ${\mathcal{O}}({M^3}{K^3})$.
		\item In the PEM ($N=M$) and the GPEM ($N<M$), where the channel vectors $\{ {{{\bf{\bar g}}}_m}\} _{m = 1}^M$ are partitioned into $N$ groups, the complexity of the Hessian matrix inversion is reduced to ${\mathcal{O}}({M^3}{K^3}/N^2)$ in each iteration.
	\end{itemize}
	\item \textbf{For random channel:}
In the case of a random channel, the E-step involves matrix inversion with a complexity of ${\mathcal{O}}({M^3}{K^3})$, as seen in \eqref{eq:co1}. In the M-step, maximizing $Q( {{\bf{\Theta }}|{{\bf{\Theta }}^{(p - 1)}}} )$ requires another Hessian matrix inversion, which also has a complexity of ${\mathcal{O}}({M^3}{K^3})$. Thus, the overall complexity of the VIEM algorithm for random channels is on the order of ${\mathcal{O}}({M^3}{K^3})$.
\end{enumerate}

As anticipated, when the number of antennas at the BS increases or a large $N$ is adopted, the GPEM and PEM algorithms exhibit significantly reduced complexities compared to the full EM algorithm in each iteration.

\section{Numerical Simulations} \label{sec:simu}

Numerical simulations were carried out to validate our theoretical findings and the effectiveness of the proposed channel estimators for  massive MIMO systems with T-ADCs.  The pilot and data were randomly drawn from the QPSK constellation unless otherwise indicated. All the elements in $\bf{H}$ were generated from the i.d.d. Gaussian distribution with zero mean and unit variance, and the noise variance was $\sigma^2 = 1$, unknown at the BS.
In the algorithm comparison part, we used the normalized CRLB (NCRLB) and the normalized mean square error (NMSE), which are respectively defined as
\begin{eqnarray}
{\rm{NCRLB}}\left( {\bf{H}} \right) \!=\! \frac{{{\rm{CRLB}}\left( {\bf{H}} \right)}}{{\left\| {\bf{H}} \right\|_{\mathrm F}^2}},{\rm{NMSE \!=\! {\mathbb{E}}}}\left[ {\frac{{\left\| {{\bf{\hat H}} \!-\! {\bf{H}}} \right\|_{\mathrm F}^2}}{{\left\| {\bf{H}} \right\|_{\mathrm F}^2}}} \right].
\end{eqnarray}
All the results were obtained by averaging over $3000$ independent realizations.

\begin{figure*}[t!]
	\subfigure[]{\includegraphics[width=0.48\linewidth]{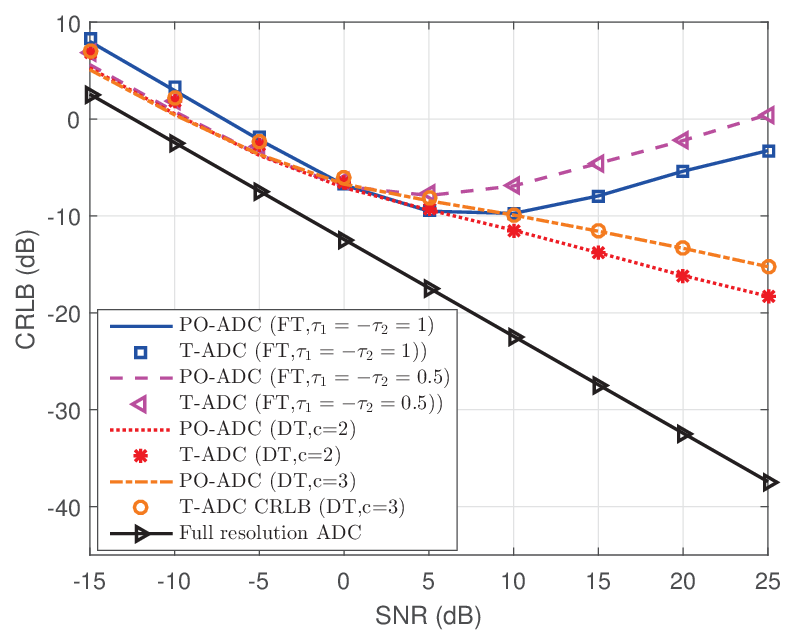} \label{2_a}}
	\subfigure[]{\includegraphics[width=0.48\linewidth]{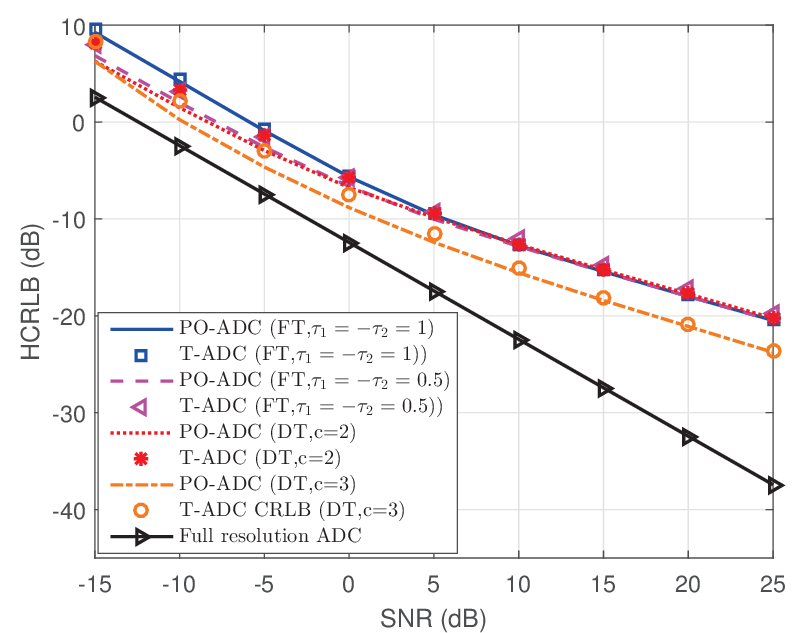}\label{2_b}}
	\hfill
	\subfigure[]{\includegraphics[width=0.48\linewidth]{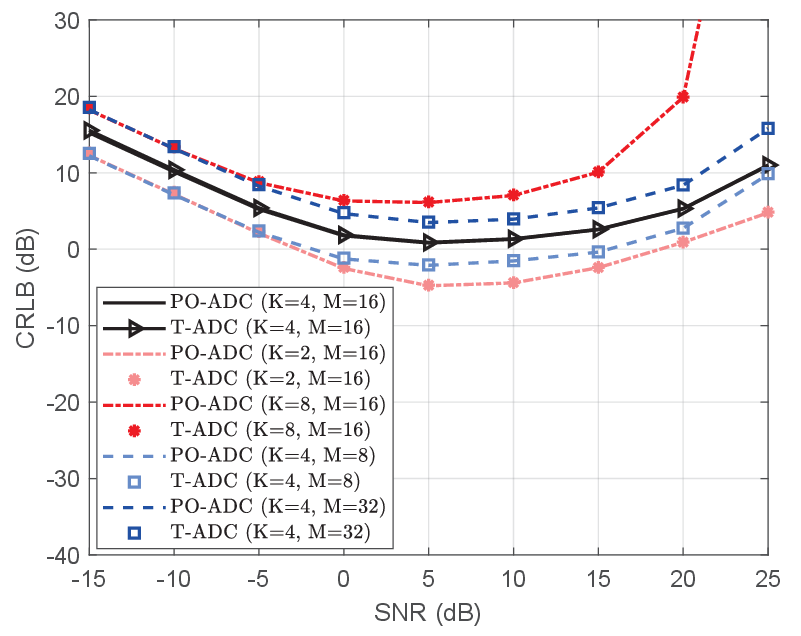}\label{2_c}}
	\subfigure[]{\includegraphics[width=0.48\linewidth]{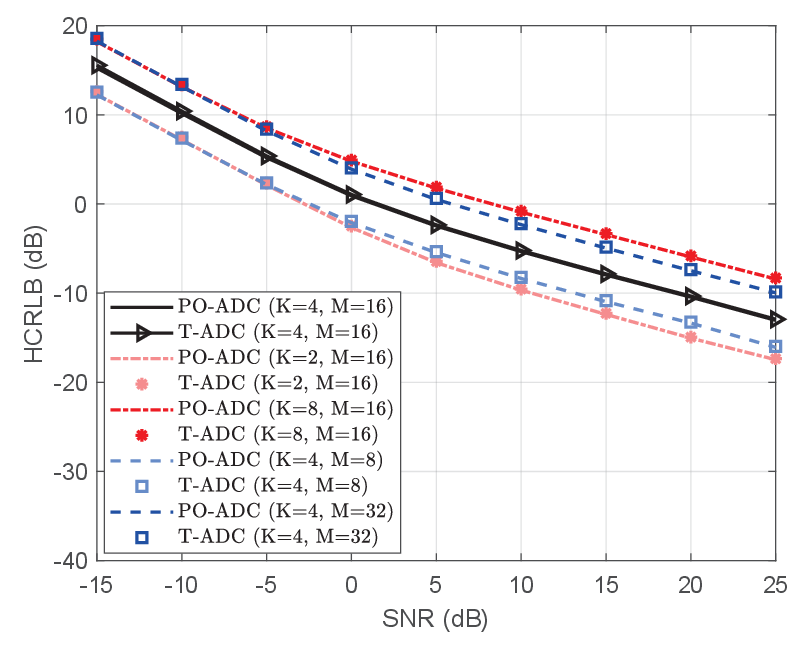}\label{2_d}}
	\caption{CRLBs of the PA estimation with FT and DT schemes as functions of SNRs in MIMO systems with two-threshold ADCs, where ${T_{\rm{p}}} = 100$. (a) Deterministic CRLB ($K=2$, $M=8$), (b) HCRLB ($K=2$, $M=8$), and (c) Deterministic CRLB with respect to $K\in\{2, 4, 8\}$, $M\in\{8, 16, 32\}$, (d) HCRLB with respect to $K\in\{2, 4, 8\}$, $M\in\{8, 16, 32\}$.}
	\label{Fig_CRLB}
\end{figure*}
\begin{figure}[t!]
	\centering
		\hspace{-2cm}
	{\begin{minipage}[h]{0.48\textwidth}
			\centering
			\epsfig{figure=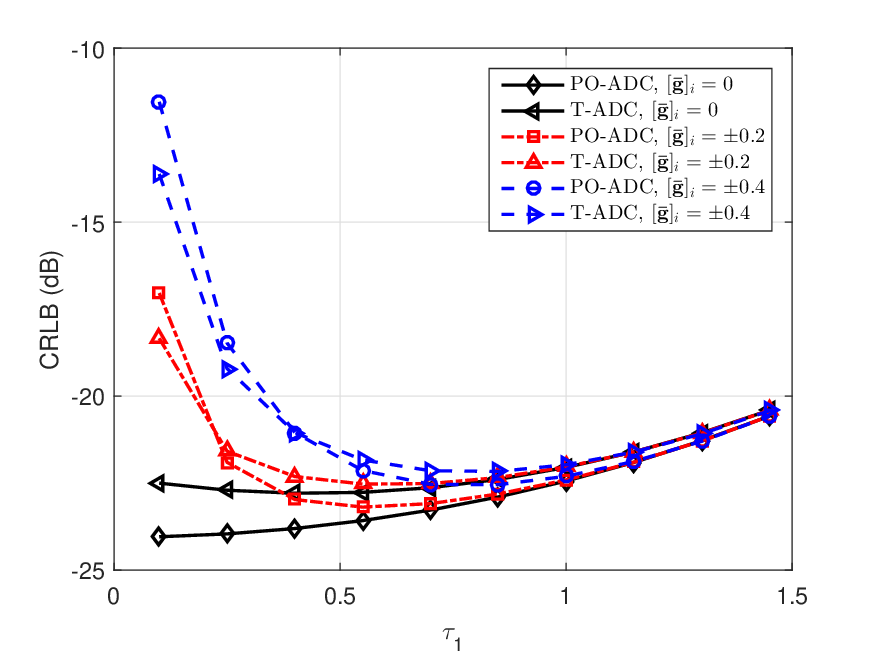,width=1.3\columnwidth}
		\end{minipage}
	}
	\hfill
	\caption{Deterministic CRLBs of the PA estimation in SISO systems with two-threshold ADCs against varying thresholds, where ${T_{\rm{p}}} = 100$, and a symmetrical threshold pair $\tau_1 = -\tau_2$ was used.}
	\label{Fig_SISO}
\end{figure}

\subsection{CRLB of the PA Estimation}
\label{sec_sim_PA}

We first examined the performance limits of the PA estimation in low-power massive MIMO systems with two-threshold ADCs. Two thresholding schemes were considered: a) Fixed thresholding (FT), where all quantizers used fixed thresholds ${\tau _1} =  -{\tau _2} = \Delta $, independent of the analog input; b) Dynamic thresholding (DT), where the thresholds were dynamically adjusted based on ${\tau _1} =  -{\tau _2} = {\mathbb{E}}[{\left\| {\bf{Y}} \right\|_{\mathrm F}}]/(cMT)$, with $c$  a positive constant and ${\rm{E}}[{\left\| {\bf{Y}} \right\|_{\mathrm F}}]$ measured by the automatic gain control unit.

As shown in Fig. \ref{Fig_CRLB}, the deterministic CRLB for the T-ADC is consistently lower than that for the PO-ADC, with the gap diminishing as the SNR increases. However, a noticeable performance gap persists in two-threshold ADC systems compared to full-resolution systems, particularly at high SNRs. Moreover, we observe that the DT scheme outperforms the FT scheme in most cases. The FT scheme exhibits the distinct behavior of decreasing CRLB followed by an increase as SNR grows. Given that noise variance is fixed at $\sigma^2=1$, lower SNR implies lower transmit power. At low transmit power, signals in the outer intervals $(\tau_1, +\infty)$ or $(-\infty, \tau_2)$ are misclassified into the middle interval $[\tau_2, \tau_1]$, while at higher transmit power, signals in the middle interval are misclassified into the outer intervals. Consequently, the lowest CRLB is observed at a moderate SNR. Furthermore, this explains why lower CRLBs occur with $\tau = 0.5$ for low SNR and $\tau = 1$ for high SNR. In contrast, the DT scheme's adaptive thresholding significantly improves accuracy in T-ADC systems. Additionally, in Fig.~\ref{Fig_CRLB} \subref{2_c} and Fig.~\ref{Fig_CRLB} \subref{2_d}, we depict the CRLB and HCRLB performance as a function of $K$ and $M$. For both deterministic and random channels, the CRLB and HCRLB increase with larger values of $M$ or $K$.

Fig. \ref{Fig_SISO} plots deterministic CRLBs in \eqref{eq:crbsp} and \eqref{eq:crbst} of SISO systems with two-threshold ADCs against varying thresholds. We observe that when the channel magnitude $|[{\bf{\bar g}}]_i|$ is getting large, the deterministic CRLB becomes more sensitive to a smaller threshold pair, e.g., the CRLB exceeds $-15\;{\rm{dB}}$ when ${{{[{\bf{\bar g}}]}_i}} = \pm0.4$ and $|\tau_1|=|\tau_2|<0.2$. For a specific channel ${{{[{\bf{\bar g}}]}_i}}$,  the deterministic CRLBs in SISO systems with PO-ADCs and T-ADCs exhibit differences against varying thresholds, depending on the relative distance between the terms ${{{[{\bf{\bar g}}]}_i}}  - {\tau _1}$ and ${{{[{\bf{\bar g}}]}_i}}  - {\tau _2}$, as evidenced by the analysis in Section \ref{sec:siso}.

\subsection{Performance of the JPD Estimation}
\label{sec_sim_HCELB}

\begin{figure}[t!]
	\hspace{-2cm}
	\centering
	{\begin{minipage}[h]{0.48\textwidth}
			\centering
			\epsfig{figure=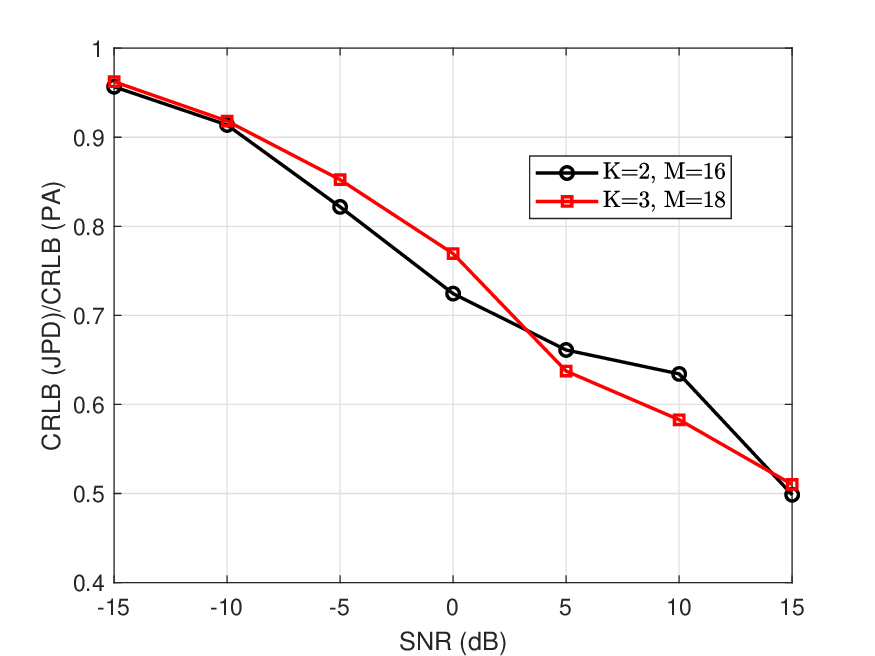,width=1.3\columnwidth}
		\end{minipage}
	}
	\hfill
	\caption{The ratio of HCRLB (JPD) to CRLB (PA) as a function of the SNR, where $T_{\rm{p}} = T_{\rm{d}} = 20$, and the DT scheme was used with $c=3$.}
	\label{Fig_CRLB_JPD_ratio}
\end{figure}

As indicated by the proof of \textit{Theorem 5} and \eqref{eq:FIMJPD}, the JPD estimation can achieve a very comparable performance as PA when the number of BS antennas $M$ is much greater than the number of users $K$, therefore, we are more concerned on its robustness when combating the negative effect caused by a small $M$. Fig. \ref{Fig_CRLB_JPD_ratio} depicts the ratio of the deterministic CRLB (JPD) to the deterministic CRLB (PA) as a function of the SNR.  Observe that the ratio is close to the unity in the low SNR region, which implies that the detected data matrix ${\bf{X}}_{\rm{d}}$ as the virtual pilot does not provide sufficient measured information to refine the channel estimate. This is because the quantized outputs corresponding to the distinct sending vectors lie into the overlapping regions which cannot be classified effectively in baseband, as analyzed in Appendix D. Moreover, when the SNR increases from $10\;{\rm{dB}}$ to $15\;{\rm{dB}}$, this ratio becomes smaller and gradually approximates to a half such that the JPD estimation achieves a nearly identical performance as the PA one. An asymptotic curve approaching 0.5 would emerge if the SNR becomes higher (exceeding 15 dB).
Both \textit{Theorem 5} and the numerical results indicate that, given a high SNR, the JPD estimation greatly saves the pilot overhead for the quantized MIMO system.

\subsection{Proposed Channel Estimators}

\begin{figure}[t!]
	\hspace{-2cm}
	\centering
	{\begin{minipage}[h]{0.5\textwidth}
			\centering
			\epsfig{figure=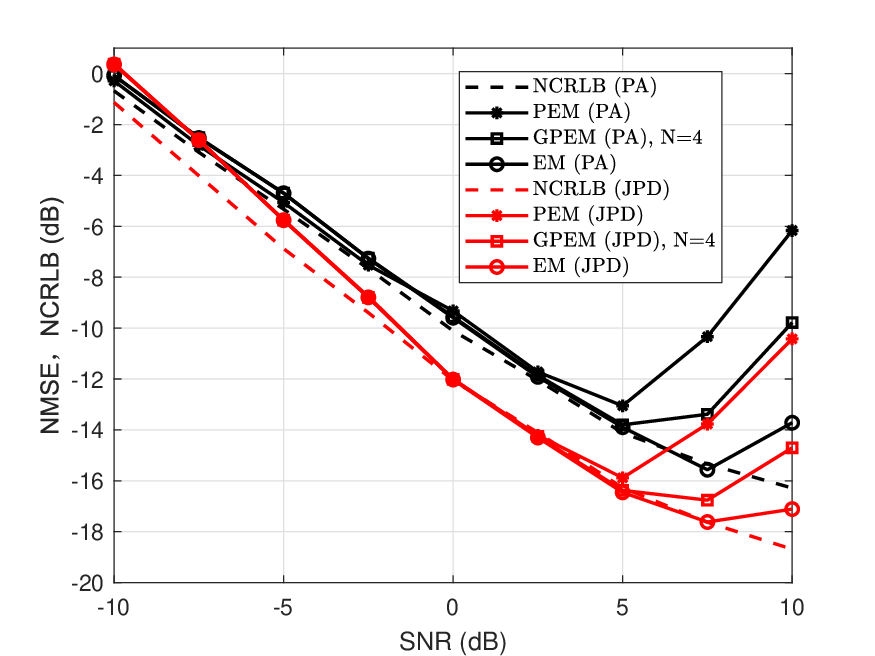,width=1.3\columnwidth}			
		\end{minipage}
	}
	\hfill
	\caption{NMSEs of the modified EM, PEM and GPEM algorithms over a deterministic channel, where $T_{\rm{p}}=T_{\rm{d}} = 30$, $M=64$, $K=4$, and the DT scheme was used with $c=3$.}
	\label{Fig_PEM_EM_TADC}
\end{figure}

We next demonstrated the effectiveness of the proposed channel estimators in energy-efficient AIoT systems based on massive MIMO with T-ADCs. Fig. \ref{Fig_PEM_EM_TADC} plots NMSEs of proposed EM based algorithms and their associated NCRLBs for the deterministic channel. It shows that the pilot enables these estimators closely attain the corresponding NCRLBs when SNR $\le 7.5\;{\rm{dB}}$, and the JPD scheme cannot even lower their NMSEs in the low SNR region, consistent with the results in Fig. \ref{Fig_CRLB_JPD_ratio}. We also find that because the coarse quantization effect is amplified in the high SNR region, all the algorithms operate inefficiently. This is also the area that the EM algorithm outperforms its PEM and GPEM counterparts. Their NMSEs become almost equivalent when the SNR is low, e.g., below $2.5\;{\rm{dB}}$, as evidenced in Appendix E.

\begin{figure}[t!]
	%	\vspace{-1cm}
	\hspace{-2cm}
	\centering
	{\begin{minipage}[h]{0.5\textwidth}
			\centering
			\epsfig{figure=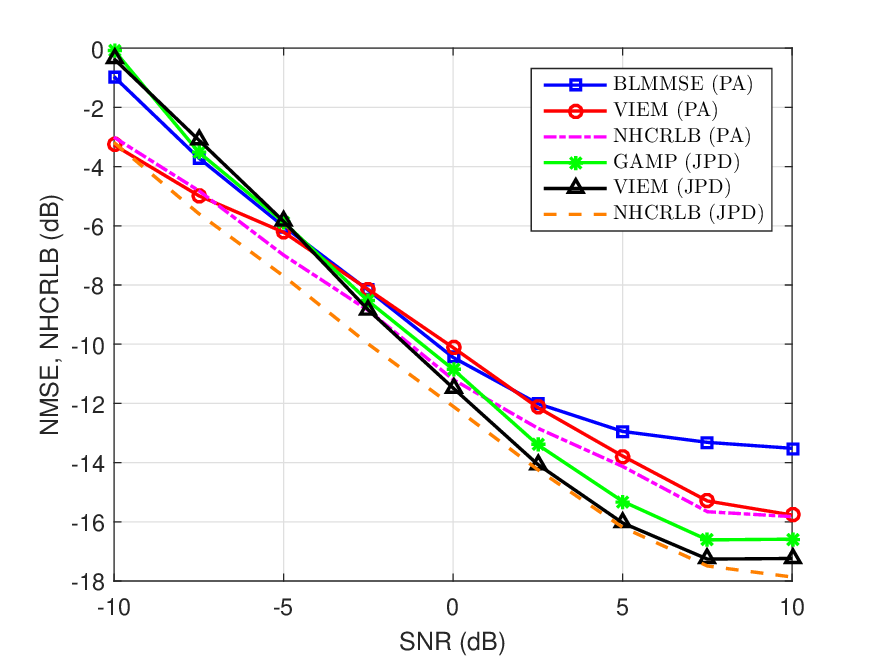,width=1.3\columnwidth}
		\end{minipage}
	}
	\hfill
	\caption{NMSEs of the proposed VIEM and state-of-the-art baseline algorithms over a random channel, where $T_{\rm{p}}=T_{\rm{d}} = 30$, $M=64$, $K=4$, and the DT scheme was used with $c=3$.}
	\label{Fig_VIEM}
\end{figure}
\begin{figure}[t!]
	\hspace{-2cm}
	\centering
	{\begin{minipage}[h]{0.5\textwidth}
			\centering
			\epsfig{figure=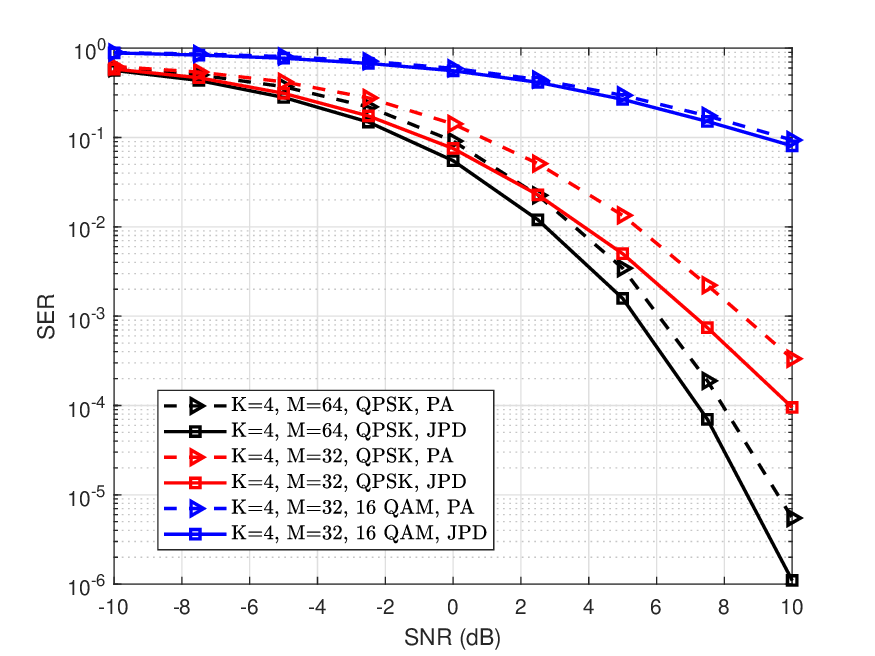,width=1.3\columnwidth}
		\end{minipage}
	}
	\hfill
	\caption{SERs of VIEM (PA) and VIEM (JPD) as functions of SNRs for random channel estimation, where $T_{\rm{p}}= 30, T_{\rm{d}} = 500$, and the DT scheme was used with $c=3$.}
	\label{Fig_SER}
\end{figure}

For the random channel, Fig. \ref{Fig_VIEM} compares NMSEs of the proposed VIEM algorithm with the associated normalized HCRLB (NHCRLB) and two state-of-the-art algorithms, i.e., the Bussgang based linear minimum mean square error estimator (BLMMSE) \cite{li2017channel} and the generalized approximate message passing (GAMP) algorithm \cite{wen2015bayes, mo2017channel, al2017gamp}. We observe that it is reasonable to exploit the detected data at the medium-to-high SNR region, where the NMSE of VIEM (JPD) is lower than those of VIEM (PA), BLMMSE (PA) and GAMP (JPD). Note that, the JPD scheme runs in a turbo fashion, so that its effectiveness highly depends on how accurate the detected data in each iteration. It becomes inefficient at the low SNR region, because the erroneous symbols, due to the heavy noise, break down the reliability of the detected data for channel enhancement. Furthermore, the efficiency comparisons are presented in Table. \ref{tab_efficiency}. Different from BLMMSE and GAMP, which aim to devise low-complexity systems, our VIEM algorithm focuses on improving channel estimation performance in low-resolution systems. Although dominated by the matrix inversion complexity $\mathcal{O}(M^3K^3)$, the runtime for PA (0.278s) and JPD (0.510s) schemes are both acceptable for practical systems.
\begin{table}[t!]
	\caption{Efficiency Comparisons of the Proposed VIEM and Baselines}
	\centering
	\begin{tabular}{c|c|c|c|c}
		\hline	
		
		\hline
		
		\hline
		Methods & BLMMSE(PA) & VIEM(PA) & GAMP(JPD) & VIEM(JPD) \\
		\hline
		
		\hline
		Computational Complexity & $\mathcal{O}(T_\mathrm{p}^2)$ & $\mathcal{O}(M^3K^3)$ & $\mathcal{O}(MK)$ & $\mathcal{O}(M^3K^3)$ \\
		\hline
		Runtime (second) & 0.003 & 0.278 & 0.002 & 0.510 \\
		\hline
		
		\hline	
		
		\hline
	\end{tabular}
	\label{tab_efficiency}
\end{table}

The symbol error rate (SER) of the VIEM as a function of the SNR is plotted in Fig. \ref{Fig_SER}. Again, the benefits of the JPD scheme become increasingly apparent at higher SNR levels. For varying numbers of UEs, denoted by $K$, the SER performance improves as the ratio $M/K$ increases. However, a significant degradation in SER is observed when employing higher-order constellations, such as 16-QAM, in comparison to QPSK. This degradation can be mitigated by increasing both the number of BS antennas, $M$, and the symbol power, $P_{\rm{s}}$. We also observe from Fig.~\ref{Fig_SER} that, increasing $M$ not only enhances channel estimation but also improves demodulation performance. This is because a larger number of antennas translates into more receiving channels. This, in turn, increases the number of quantization intervals. As a result, quantization noise diminishes, leading to an increase in the estimation and detection SNR. On the other hand, the performance of the JPD scheme tends to decline as modulation orders rise. In low-resolution systems, finite bit quantization introduces greater quantization noise during decoding, which reduces the system's ability to effectively demodulate higher-order signals, thereby limiting the demodulation freedom of the JPD system.

Finally, the computational time of all the proposed EM based algorithms is provided in Table \ref{tab_1}, over 100 realizations implemented at an Intel core i7 machine with $2.8\;{\rm{GHz}}$ processor and $8\;{\rm{GB}}$ RAM. For the deterministic channel, the much heavy computational burden of EM can be observed, as compared with PEM and GPEM. This is in accordance with the complexity analysis in Section \ref{sec:JPD}. Therefore, in practice, one can use PEM at low and medium SNR regions and EM/GPEM at the high SNR region to trade off the estimation accuracy and the computational cost. While for the random channel, VIEM requires less iterations than  EM/PEM/GPEM, and hence its computational time is acceptable, among all the proposed EM based algorithms.
\begin{table}[t!]
	\caption{Computational Time in Seconds of the Proposed EM based Algorithms for $100$ Realizations, where $T_{\rm{p}} = T_{\rm{d}} = 30$.}
	\centering
	\begin{tabular}{c|c|c|c|c}
		\hline
		
		\hline
		
		\hline
		
		Time (s) & EM & PEM & GPEM ($N=2$) & {VIEM} \\
		\hline
		
		\hline
		$K=4,M=32$ & 19.61 & 0.53 & 4.29 & {1.89} \\
		\hline
		$K=4,M=64$ & 51.37 & 2.72 & 11.39 & {7.72} \\
		\hline
		$K=6,M=64$ & 70.18 & 4.51 & 18.56 & {12.68}\\
		
		\hline
		
		\hline
		
		\hline
	\end{tabular}
	\label{tab_1}
\end{table}

\section{Conclusion}
\label{sec:conclu}

This work has considered the contact-free smart sensing in AIoT and has investigated a low-power massive MIMO scheme using T-ADCs. We have first introduced a PO-ADC architecture to characterize the performance limit of two-threshold ADCs. For the PA estimation, the CRLBs have been derived for both deterministic and random channels, and we have proven that there exists at least $\uppi/2$ MSE loss, compared with the full resolution ADC system. To attain an enhanced channel estimate without sacrificing too many  pilot symbols, a JPD estimation scheme has been proposed and the associated CRLB analysis has shown that this scheme can effectively lower the MSE at the high SNR region. We have next proposed a modified EM algorithm for the deterministic channel estimation, as well as its parallel realizations. For the random channel, we have developed the VIEM algorithm based on the variational inference technique. Numerical simulations have confirmed the theoretical findings and have shown that the proposed algorithms perform sufficiently close to the derived CRLBs over a wide SNR range. We believe this paper has demonstrated the potential of using T-ADCs in quantized massive MIMO systems that enables ubiquitous and extreme AIoT connectivity for smart sensing.

\begin{acks}
This work was supported in part by the National Science and Technology Major Project under Grant No.\ 2024ZD1300200 and the Fundamental Research Funds for the Central Universities under Grant No. 2242023R40005.
\end{acks}

\bibliographystyle{ACM-Reference-Format}
\bibliography{reference}

\appendix

\section{Detailed Proof of \textit{Theorem 1}}

The gradient of the LLF in \eqref{eq:likli} with respect to $\bm \eta$ is calculated as
\begin{eqnarray}
{\nabla _{\bm{\eta }}}{{L_{\rm{p}}}}\!\left( {{\bf{\bar Z}};{\bm{\eta }}} \right) \!\!\!\!\!&=&\!\!\!\!\! \sum\limits_{i = 1}^{2{T_p}} {\sum\limits_{j = 1}^2 {\frac{{ {\left(2[{\bf{\bar Z}}]_{i,j}-1\right)}q\!\left( {{{[{\bf{A}}]}_{i,j}}} \right)}}{{Q\!\left( {{\left(2[{\bf{\bar Z}}]_{i,j}-1\right)}{{[{\bf{A}}]}_{i,j}}} \right)}}\frac{{\partial {{[{\bf{A}}]}_{i,j}}}}{{\partial {\bm{\eta }}}}} },
\end{eqnarray}
where
\begin{eqnarray}
\frac{{\partial {{[{\bf{A}}]}_{i,j}}}}{{\partial {\bm{\eta }}}} = {\left[ {\frac{{\sqrt 2 {{[{\bf{\bar X}}]}_{i,:}}}}{\sigma },\frac{{\sqrt 2 \left( {{\tau _j} - {{[{\bf{ \bar X\bar g}}]}_i}} \right)}}{{{\sigma ^2}}}} \right]^{\mathsf T}}\in\mathbb{R} {^{(2K + 1) \times 1}}.
\label{eq:gradient}
\end{eqnarray}
According to \eqref{eq:FIM}, the FIM ${\bf{F}}_{\rm{p}}({\bm{\eta }})$ can be expressed as
\begin{eqnarray}
{\bf{F}}_{\rm{p}}\!\left( {\bm{\eta }} \right) \!\!\!\!\!&=&\!\!\!\!\! {{\mathbb{E}}_{{\bf{\bar Z}}|{\bm{\eta }}}}\left[ {{\nabla _{\bm{\eta }}}{{L_{\rm p}}}\!\left( {{\bf{\bar Z}};{\bm{\eta }}} \right){{\left( {{\nabla _{\bm{\eta }}}{{L_{\rm p}}}\!\left( {{\bf{\bar Z}};{\bm{\eta }}} \right)} \right)}^{\mathsf T}}} \right] \nonumber \\
\!\!\!\!\!&\mathop  = \limits^{\rm{(a)}}&\!\!\!\!\! \sum\limits_{i = 1}^{2{T_{\rm{p}}}} {\sum\limits_{j = 1}^2 {\frac{{{q^2}({{[{\bf{A}}]}_{i,j}})}}{{Q({{[{\bf{A}}]}_{i,j}})Q( - {{[{\bf{A}}]}_{i,j}})}}\frac{{\partial {{[{\bf{A}}]}_{i,j}}}}{{\partial {\bm{\eta }}}}} {{\left( {\frac{{\partial {{[{\bf{A}}]}_{i,j}}}}{{\partial {\bm{\eta }}}}} \right)}^{\mathsf T}}}  \label{eq:fiser} \\
	\!\!\!\!\!&=&\!\!\!\!\! {\bf{M\Lambda }}{{\bf{M}}^{\mathsf T}},
	\label{eq:fiser1}
\end{eqnarray}
where (a) stems from the fact that ${{\mathbb{E}}_{{\bf{\bar Z}}|{\bm{\eta }}}}\left[ {{\nabla _{\bm{\eta }}}{{L_{\rm p}}}\!\left( {{\bf{\bar Z}};{\bm{\eta }}} \right)} \right] = {\bf{0}}$, the matrix ${\bf{M}} \in\mathbb{R} {^{(2K + 1) \times 4{T_{\rm{p}}}}}$ is with
${[{{\bf{M}}}]_{:,2(j - 1){T_{\rm{p}}} + i}} = \partial {[{\bf{A}}]_{i,j}}/\partial {\bm{\eta }}$, where $i = 1,2, \cdots, 2{T_{\rm{p}}}$, and $j = 1,2$, and the matrix ${\bf{\Lambda }} \in {{\mathbb{R}}^{4{T_{\rm{p}}} \times 4{T_{\rm{p}}}}}$ is diagonal with the elements
\begin{eqnarray*}
	{[{{\bf{\Lambda }}}]_{2(j - 1){T_{\rm{p}}} + i,2(j - 1){T_{\rm{p}}} + i}}
	= \frac{{{q^2}({{[{{\bf{A}}}]_{i,j}}})}}{{Q({{[{{\bf{A}}}]_{i,j}}})Q( - {{[{{\bf{A}}}]_{i,j}}})}},i = 1,2, \cdots ,{2T_{\rm{p}}},j = 1,2.
\end{eqnarray*}
The CRLB is the inverse of its FIM and the MSE of ${ { {\bf \hat{\bar g}}}}$ is asymptotically equal to the trace of the submatrix ${{\bf{[CRLB(}}{\bm{\eta }}{\rm{)]}}_{1:2K,1:2K}}$, which gives \eqref{eq:mse1}.

\section{Detailed Proof of \textit{Theorem 2}}

Observe that in \eqref{eq:block},  ${{\bf{F}}_{\rm{p}}}({\bf{\bar g}})$  is exactly the FIM when the noise variance $\sigma^2$ is known at the BS. In this case, we have
	\begin{eqnarray}
	\mathop {{\rm{min}}}\limits_{{\bf{\bar X}},{\tau _1},{\tau _2}} {\rm{mse_{PA,p}}}\left({ { {\bf \hat{\bar g}}}}\right) \mathop  = \limits^{{\rm{a}}{\rm{.e}}{\rm{.}}} \mathop {{\rm{min}}}\limits_{{\bf{\bar X}},{\tau _1},{\tau _2}} {\kern 1pt} {\kern 1pt} {\kern 1pt} {\rm{tr}}\left( {{{\left({{\bf{F}}_{\rm{p}}({\bf{\bar g}})} \right)}^{ - 1}}} \right) = \mathop {{\rm{min}}}\limits_{{\bf{\bar X}},{\tau _1},{\tau _2}} \sum\limits_{i = 1}^{2K} {\frac{1}{{{\lambda _i}}}},
	\label{eq:mintr}
	\end{eqnarray}
where $\{ {\tau _i}\} _{i = 1}^2$ are the thresholds of PO-ADCs, $\{ {\lambda _i}\} _{i = 1}^{2K}$ are the eigenvalues of ${{\bf{F}}_{\rm{p}}({\bf{\bar g}})}$. Using the properties of convex function $\frac{1}{\lambda_i}$, we readily obtain
\begin{equation}
	\mathop {{\rm{min}}}\limits_{{\bf{\bar X}},{\tau _1},{\tau _2}} \sum\limits_{i = 1}^{2K} {\frac{1}{{{\lambda _i}}}}
	\geq \mathop {{\rm{min}}}\limits_{{\bf{\bar X}},{\tau _1},{\tau _2}} \frac{4K^2}{\sum_{i = 1}^{2K} \lambda_i} = \frac{4K^2}{{\rm{tr}}\left( {{{\bf{F}}_{\rm{p}}}({\bf{\bar g}})} \right)}.
	\label{eq:tr1}
\end{equation}
The maximum of ${\rm{tr}}\left( {{\bf{F}}_{\rm{p}}({\bf{\bar g}})} \right)$ in \eqref{eq:tr1} is given by
	\begin{eqnarray}
	{\rm{\mathop {{\rm{max}}}\limits_{{\bf{\bar X}},{\tau _1},{\tau _2}}  tr}}\left( {{\bf{F}}_{\rm{p}}({\bf{\bar g}})} \right) \!\!\!\!\!&=&\!\!\!\!\! \mathop {{\rm{max}}}\limits_{{\bf{\bar X}},{\tau _1},{\tau _2}} {\rm{tr}}\left( {{{[{\bf{M}}]}_{1:2K,:}}{\bf{\Lambda }}{{[{{\bf{M}}^{\mathsf T}}]}_{:,1:2K}}} \right) = {\rm{  \mathop {{\rm{max}}}\limits_{{\bf{\bar X}},{\tau _1},{\tau _2}} tr}}\left( {{\bf{\Lambda }}{{[{{\bf{M}}^{\mathsf T}}]}_{:,1:2K}}{{[{\bf{M}}]}_{1:2K,:}}} \right) \nonumber \\
	\!\!\!\!\!&\mathop  = \limits^{{\rm{(a)}}} &\!\!\!\!\! \frac{{16{P_{\rm{s}}}{T_{\rm{p}}}K}}{{\uppi {\sigma ^2}}},
	\label{eq:optt}
	\end{eqnarray}
where (a) is achievable when all the elements of the matrix ${\bf{A}}$ in \eqref{eq:dA} are equal to zero such that ${[{\bf{\Lambda }}]_{i,i}} = {{q^2}(0)/{Q^2}(0)} = 2/\uppi ,\forall i$. This occurs when all the thresholds are exactly identical to the analog input both temporally and spatially, i.e., ${[{\bf{\bar X\bar g}}]_i} = {\tau _j},\forall i,j$. Then based on \eqref{eq:mintr}, \eqref{eq:tr1} and \eqref{eq:optt}, the minimum MSE of the channel estimate becomes
\begin{eqnarray}
\mathop {{\rm{min}}}\limits_{{\bf{\bar X}},{\tau _1},{\tau _2}} {\rm{mse_{PA,p}}}\left({ { {\bf \hat{\bar g}}}}\right) \ge \frac{4K^2}{{\rm{tr}}\left( {{{\bf{F}}_{\rm{p}}}({\bf{\bar g}})} \right)} = \frac{{K\uppi {\sigma ^2}}}{{4{P_{\rm{s}}}{T_{\rm{p}}}}}.
\label{eq:lower_mse}
\end{eqnarray}

Note from the quantization rule in \eqref{eq:poadc} and PO-ADC architecture in Fig. \ref{Fig:1}, a noise corrupted signal $u$ is sampled by two one-bit ADCs. If the two one-bit ADCs are substituted by two full resolution ones, given the same pilot matrix ${\bf{X}} \in {\mathbb{C}^{K \times T_{\rm{p}}}}$ (${\bf{X}}{{\bf{X}}^{\mathsf H}} = {T_{\rm{p}}}{P_{\rm{s}}}{{\bf{I}}_K}$), the MSE of the ML channel estimate ${\bf \hat{\bar g}}$ in the full resolution ADC system becomes ${\rm{mse}}_{{\rm{f}}}\left({\bf \hat{\bar g}}\right) \mathop  = \limits^{{\rm{a}}{\rm{.e}}{\rm{.}}}  {\sigma ^2}{\rm{tr((}}{\bf{X}}{{\bf{X}}^{\mathsf H}}{{\rm{)}}^{ - 1}}{\rm{)/2}} =  K{\sigma ^2}/2{P_{\rm{s}}}{T_{\rm{p}}}$ \cite[Chapter 3]{kay1993fundamentals}, \cite{stein2018performance}. Then, based upon \eqref{eq:lower_mse} that $\mathop {{\rm{min}}}\limits_{{\bf{\bar X}},{\tau _1},{\tau _2}} {\rm{mse_{PA,p}}}\left({ { {\bf \hat{\bar g}}}}\right) \ge K\uppi {\sigma ^2}/4{P_{\rm{s}}}{T_{\rm{p}}}$, we have \eqref{eq:lower_mse1}.

\section{Detailed Proof of \textit{Theorem 4}}

By differentiating the LLFs with respect to ${{\bm{\theta }}_i}$ and letting the gradient equal to zero, i.e.,
\begin{eqnarray}
{\nabla _{{{\bm{\theta }}_i}}}{L_{\rm{p}}}\!\left( {{\bf{\bar Z}};{\bm{\eta }}} \right) = \left[ {\begin{array}{*{20}{c}}
	{\sum\limits_{j = 1}^{{T_{\rm{p}}}} {\frac{{q\left( {{{[{{\bm{\theta }}_i}]}_1}} \right)\left( {2{{[{\bf{\bar Z}}]}_{(i - 1){T_{\rm{p}}} + j,1}} - 1} \right)}}{{Q\left( {\left( {2{{[{\bf{\bar Z}}]}_{(i - 1){T_{\rm{p}}} + j,1}} - 1} \right){{[{{\bm{\theta }}_i}]}_1}} \right)}}} }\\
	{\sum\limits_{j = 1}^{{T_{\rm{p}}}} {\frac{{ {q\left( {{{[{{\bm{\theta }}_i}]}_2}} \right)\left( {2{{[{\bf{\bar Z}}]}_{(i - 1){T_{\rm{p}}} + j,2}} - 1} \right)} }}{{Q\left( {\left( {2{{[{\bf{\bar Z}}]}_{(i - 1){T_{\rm{p}}} + j,2}} - 1} \right){{[{{\bm{\theta }}_i}]}_2}} \right)}}} }
	\end{array}} \right] = {\bf{0}},
\end{eqnarray}
\begin{eqnarray}
{\nabla _{{{\bm{\theta }}_i}}}{L_{\rm{t}}}\!\left( {{\bf{\bar Z}};{\bm{\eta }}} \right) = \left[ {\begin{array}{*{20}{c}}
	{\frac{{\sum\limits_{j = 1}^{{T_{\rm{p}}}} {q\left( {{{[{{\bm{\theta }}_i}]}_1}} \right){{[{\bf{\bar Z}}]}_{(i - 1){T_{\rm{p}}} + j,1}}} }}{{Q\left( {{{[{{\bm{\theta }}_i}]}_1}} \right)}} - \frac{{\sum\limits_{j = 1}^{{T_{\rm{p}}}} {q\left( {{{[{{\bm{\theta }}_i}]}_1}} \right)\left( {{{[{\bf{\bar Z}}]}_{(i - 1){T_{\rm{p}}} + j,2}} - {{[{\bf{\bar Z}}]}_{(i - 1){T_{\rm{p}}} + j,1}}} \right)} }}{{Q\left( {{{[{{\bm{\theta }}_i}]}_2}} \right) - Q\left( {{{[{{\bm{\theta }}_i}]}_1}} \right)}}}\\
	{\frac{{\sum\limits_{j = 1}^{{T_{\rm{p}}}} {q\left( {{{[{{\bm{\theta }}_i}]}_2}} \right)\left( {{{[{\bf{\bar Z}}]}_{(i - 1){T_{\rm{p}}} + j,2}} - 1} \right)} }}{{1 - Q\left( {{{[{{\bm{\theta }}_i}]}_2}} \right)}} + \frac{{\sum\limits_{j = 1}^{{T_{\rm{p}}}} {q\left( {{{[{{\bm{\theta }}_i}]}_2}} \right)\left( {{{[{\bf{\bar Z}}]}_{(i - 1){T_{\rm{p}}} + j,2}} - {{[{\bf{\bar Z}}]}_{(i - 1){T_{\rm{p}}} + j,1}}} \right)} }}{{Q\left( {{{[{{\bm{\theta }}_i}]}_2}} \right) - Q\left( {{{[{{\bm{\theta }}_i}]}_1}} \right)}}}
	\end{array}} \right] = {\bf{0}},
\end{eqnarray}
we can prove that the ML estimate on ${{\bm{\theta }}_i}$ is given by
\begin{eqnarray}
{{{\bm{\hat \theta }}}_i} = \left[ {\begin{array}{*{20}{c}}
	{{Q^{ - 1}}\left( {\frac{1}{{{T_{\rm{p}}}}}\sum\limits_{j = 1}^{{T_{\rm{p}}}} {{{{[{\bf{\bar Z}}]}_{(i - 1){T_{\rm{p}}} + j,1}}}} } \right)}\\
	{{Q^{ - 1}}\left( {\frac{1}{{{T_{\rm{p}}}}}\sum\limits_{j = 1}^{{T_{\rm{p}}}} {{{{[{\bf{\bar Z}}]}_{(i - 1){T_{\rm{p}}} + j,2}}}} } \right)}
	\end{array}} \right],i=1,2,
\label{eq:MLEt}
\end{eqnarray}
which holds for SISO systems with both PO-ADCs and T-ADCs. Based on the invariance property \cite{ribeiro2006bandwidth}, the ML channel estimate $[{\bf\hat{\bar g}}]_i$ can be obtained as
\begin{eqnarray}
{[{\bf \hat{ \bar g}}]_i} = \frac{1}{{2\sqrt {{P_{\rm{s}}}} }} \left({( - 1)^{i - 1}}\frac{{{{[{{{\bm{\hat \theta }}}_1}]}_1}{\tau _2} - {{[{{{\bm{\hat \theta }}}_1}]}_2}{\tau _1}}}{{{{[{{{\bm{\hat \theta }}}_1}]}_1} - {{[{{{\bm{\hat \theta }}}_1}]}_2}}}\right. + \left. \frac{{{{[{{{\bm{\hat \theta }}}_2}]}_1}{\tau _2} - {{[{{{\bm{\hat \theta }}}_2}]}_2}{\tau _1}}}{{{{[{{{\bm{\hat \theta }}}_2}]}_1} - {{[{{{\bm{\hat \theta }}}_2}]}_2}}}\right),i = 1,2.
\label{eq:inp}
\end{eqnarray}

We next derive the deterministic CRLB of SISO systems with two-threshold ADCs. Similar to the analysis in \textit{Theorem 1}, for the intermediate parameter ${{{\bm{\theta }}_i}}$, the FIMs in PO-ADCs and T-ADCs are respectively given by
\begin{eqnarray}
{{\bf{F}}_{\rm{p}}}\!\left( {{{\bm{\theta }}_i}} \right) = {T_{\rm{p}}}\left[ {\begin{array}{*{20}{c}}
	{\frac{{{q^2}({{[{{{\bm{ \theta }}}_i}]}_1})}}{{Q({{[{{{\bm{ \theta }}}_i}]}_1})Q( - {{[{{{\bm{ \theta }}}_i}]}_1})}}}&{0}\\
	{0}&{\frac{{{q^2}({{[{{{\bm{ \theta }}}_i}]}_2})}}{{Q({{[{{{\bm{ \theta }}}_i}]}_2})Q( - {{[{{{\bm{ \theta }}}_i}]}_2})}}}
	\end{array}} \right],\\
{{\bf{F}}_{\rm{t}}}\!\left( {{{\bm{\theta }}_i}} \right) =
{T_{\rm{p}}}\left[ {\begin{array}{*{20}{c}}
	\!\!\!{\frac{{Q({{[{{\bm{\theta }}_i}]}_2}){q^2}({{[{{\bm{\theta }}_i}]}_1})}}{{Q({{[{{\bm{\theta }}_i}]}_1})\left( {Q({{[{{\bm{\theta }}_i}]}_2}) - Q({{[{{\bm{\theta }}_i}]}_1})} \right)}}}&\!\!\!\!\!\!\!\!\!\!\!\!\!{\frac{{ - q({{[{{\bm{\theta }}_i}]}_1})q({{[{{\bm{\theta }}_i}]}_2})}}{{{\rm{Q}}({{[{{\bm{\theta }}_i}]}_2}) - {\rm{Q}}({{[{{\bm{\theta }}_i}]}_1})}}}\\
	{\frac{{ - q({{[{{\bm{\theta }}_i}]}_1})q({{[{{\bm{\theta }}_i}]}_2})}}{{Q({{[{{\bm{\theta }}_i}]}_2}) - Q({{[{{\bm{\theta }}_i}]}_1})}}}&\!\!\!\!\!\!\!\!\!\!\!{\frac{{Q( - {{[{{\bm{\theta}}_i}]}_1}){q^2}({{[{{\bm{\theta }}_i}]}_2})}}{{Q( - {{[{{\bm{\theta }}_i}]}_2})\left( {Q({{[{{\bm{\theta }}_i}]}_2}) - Q({{[{{\bm{\theta }}_i}]}_1})} \right)}}}\!\!\!\!
	\end{array}} \right].~~~
\end{eqnarray}
The ${{\bf{F}}_{\rm{t}}}\!\left( {{{\bm{\theta }}_i}} \right)$ can also be derived from the special case of Eq. \eqref{eq:fimt} with $M=K=1$, where $\boldsymbol{\theta}_i = \left[[\mathbf{A}]_{i,1}, [\mathbf{A}]_{i,2}\right]$.
Based upon the relationship in \eqref{eq:inp} and vector transformation lemma in \cite[p. 45]{kay1993fundamentals}, we have
\begin{eqnarray}
{\rm{CRL}}{{\rm{B}}_{\rm{p/t}}}\left( {{{[{\bf{\bar g}}]}_i}} \right)
\!\!\!\!\!&\mathop  = \limits^{} &\!\!\!\!\! \sum\limits_{j = 1}^2 {\left[ {\frac{{\partial {{[{\bf{\bar g}}]}_i}}}{{\partial {{[{{\bm{\theta }}_j}]}_1}}},\frac{{\partial {{[{\bf{\bar g}}]}_i}}}{{\partial {{[{{\bm{\theta }}_j}]}_2}}}} \right]{{\left( {{{\bf{F}}_{\rm{p/t}}}\!\left( {{{\bm{\theta }}_j}} \right)} \right)}^{ - 1}}} {\left[ {\frac{{\partial {{[{\bf{\bar g}}]}_i}}}{{\partial {{[{{\bm{\theta }}_j}]}_1}}},\frac{{\partial {{[{\bf{\bar g}}]}_i}}}{{\partial {{[{{\bm{\theta }}_j}]}_2}}}} \right]^{\mathsf T}} ,i=1,2.
\end{eqnarray}
After a few mathematical manipulations, we arrive at \eqref{eq:crbsp} and \eqref{eq:crbst}.

\section{Detailed Proof of \textit{Theorem 5}}

Since both the PO-ADC and the T-ADC have two nonidentical thresholds, each quantized output may lie in one of the three possible intervals. Therefore, for $M$ receiving antennas, there exists ${3^{M}}$ possible quantized regions corresponding to a real-valued analog input. Suppose that the quantized outputs associated with the noise-free input signal ${\bf{H}}{{\bf{x}}_l},l=1,2,\cdots,L,$ are equally likely to distribute in the ${3^{M}}$ quantized regions . When ${3^{M}} \gg {\mathcal{|S|}}{^K} = L$, the probability that all the possible distinct quantized outputs belong to different quantized regions can be calculated as
\begin{eqnarray}
{\rm{P}}{{\rm{r}}_{\rm{d}}} = {\prod\limits_{c = 0}^{{{\mathcal{|S|}}^K}} {\frac{{{3^{M}} - c + 1}}{{{3^{M}}}} > \left( {1 - \frac{{{{\mathcal{|S|}}^K} - 1}}{{{3^{M}}}}} \right)} ^{{{\mathcal{|S|}}^K}}},
\label{eq:prd}
\end{eqnarray}
which approaches to unity. {At the high SNR region,} the probability ${\rm{Pr}}({[{{{\bf{\bar Z}}}_{\rm{d}}}]_i},{\bm{\chi }}|{[{{\bf{X}}_{\rm{d}}}]_{:,i}} = {{\bf{x}}_{{l_1}}}) \ne 0$ implies that ${\rm{Pr}}({[{{{\bf{\bar Z}}}_{\rm{d}}}]_i},{\bm{\chi }}|{[{{\bf{X}}_{\rm{d}}}]_{:,i}} = {{\bf{x}}_{{l_2}}}) \approx 0, {l_1} \ne {l_2}$, i.e., all the quantized outputs of ${\bf{H}}{{\bf{x}}_l} + {\bf{N}},l=1,2,\ldots,L$ lie in the non-overlapping quantized regions. Using this fact, the FIM of the NPA estimation can be expressed as
\begin{eqnarray}
{\bf{F}}{_{{\rm{NPA}}}}({\bm{\chi }}) \!\!\!\!\!&=&\!\!\!\!\! {{\mathbb{E}}_{{\bf{\bar Z}}_{\rm{d}}|{\bm{\chi }}}}\left[ {{\nabla _{\bm{\chi }}}{{L_{\rm p/t}}}( {{\bf{\bar Z}}_{\rm{d}};{\bm{\chi}}} ){{\left( {{\nabla _{\bm{\chi }}}{{L_{\rm p/t}}}( {{\bf{\bar Z}}_{\rm{d}};{\bm{\chi }}} )} \right)}^{\mathsf T}}} \right] \nonumber\\
\!\!\!\!\!&=&\!\!\!\!\! {{\mathbb{E}}_{{{{\bf{\bar Z}}}_{\rm{d}}}|{\bm{\chi }}}} \left[ \frac{{\partial \ln \prod\limits_{i = 1}^{{T_{\rm{d}}}} {\sum\limits_{l = 1}^L {\Pr \left( {\left. {{{{\bf{\bar z}}}_i},{\bm{\chi }}} \right|{{\bf{x}}_l}} \right)\Pr \left( {{{[{{\bf{X}}_{\rm{d}}}]}_{:,i}} = {{\bf{x}}_l}} \right)} } }}{{\partial {\bm{\chi }}}}\right. \nonumber \\\!\!\!\!\!&\cdot&\!\!\!\!\!\left. {{\left( {\frac{{\partial \ln \prod\limits_{i = 1}^{{T_{\rm{d}}}} {\sum\limits_{l = 1}^L {\Pr \left( {\left. {{{{\bf{\bar z}}}_i},{\bm{\chi }}} \right|{{\bf{x}}_l}} \right)\Pr \left( {{{[{{\bf{X}}_{\rm{d}}}]}_{:,i}} = {{\bf{x}}_l}} \right)} } }}{{\partial {\bm{\chi }}}}} \right)}^{\mathsf T}} \right]\nonumber\\
\!\!\!\!\!&=&\!\!\!\!\! \sum\limits_{i = 1}^{{T_{\rm{d}}}} {{{\mathbb{E}}_{{{{\bf{\bar z}}}_i}|{\bm{\chi }}}}\left[ {\frac{{\sum\limits_{l = 1}^L {\frac{{\partial {\rm{Pr}}({{{\bf{\bar z}}}_i},{\bm{\chi }}|{{\bf{x}}_l})}}{{\partial {\bm{\chi }}}}} }}{{\sum\limits_{l = 1}^L {{\rm{Pr}}({{{\bf{\bar z}}}_i},{\bm{\chi }}|{{\bf{x}}_l})} }}{{\left( {\frac{{\sum\limits_{l = 1}^L {\frac{{\partial {\rm{Pr}}({{{\bf{\bar z}}}_i},{\bm{\chi }}|{{\bf{x}}_l})}}{{\partial {\bm{\chi }}}}} }}{{\sum\limits_{l = 1}^L {{\rm{Pr}}({{{\bf{\bar z}}}_i},{\bm{\chi }}|{{\bf{x}}_l})} }}} \right)}^{\mathsf T}}} \right]}  \nonumber\\
\!\!\!\!\!&=&\!\!\!\!\! \sum\limits_{i = 1}^{{T_{\rm{d}}}} {{{\mathbb{E}}_{{{{\bf{\bar z}}}_i}|{\bm{\chi }}}}\left[ {\frac{{\frac{{\partial {\rm{Pr}}({{{\bf{\bar z}}}_i},{\bm{\chi }})}}{{\partial {\bm{\chi }}}}}}{{{\rm{Pr}}({{{\bf{\bar z}}}_i},{\bm{\chi }})}}{{\left( {\frac{{\frac{{\partial {\rm{Pr}}({{{\bf{\bar z}}}_i},{\bm{\chi }})}}{{\partial {\bm{\chi }}}}}}{{{\rm{Pr}}({{{\bf{\bar z}}}_i},{\bm{\chi }})}}} \right)}^{\mathsf T}}} \right]}  \nonumber\\
\!\!\!\!\!&=&\!\!\!\!\!\sum\limits_{i = 1}^{{T_{\rm{d}}}} {{{\mathbb{E}}_{{{{\bf{\bar z}}}_i}|\bm\chi }}\left[ {\frac{{\partial \ln {\rm{Pr}}\left( {{{{\bf{\bar z}}}_i},{\bm{\chi }}} \right)}}{{\partial {\bm{\chi }}}}{{\left( {\frac{{\partial \ln {\rm{Pr}}\left( {{{{\bf{\bar z}}}_i},{\bm{\chi }}} \right)}}{{\partial {\bm{\chi }}}}} \right)}^{\mathsf T}}} \right]}  \nonumber\\
\!\!\!\!\!&\mathop  = \limits^{({\rm{a}})} &\!\!\!\!\! {{\bf{F}}_{{\rm{PA}}}}({\bm{\chi }}),
\label{eq:equivalence}
\end{eqnarray}
where (a) is valid when ${{\bf{X}}_{\rm{p}}} = {{\bf{X}}_{\rm{d}}}$.

\section{MSE Performance of the Proposed PEM Algorithm}

Recall \eqref{eq:blockF} and \eqref{eq:block}, the MSE of PEM is given by
\begin{eqnarray}
{\rm{ms}}{{\rm{e}}_{{\rm{PEM}}}}({\bf{\hat H}}) \!\!\!\!\!&=&\!\!\!\!\!\! \sum\limits_{m = 1}^M {{\rm{ms}}{{\rm{e}}_{{\rm{PEM}}}}({{ {{\bf \hat{\bar g}} }}_m})}
\mathop  = \limits^{{\rm{a.e.}}} \sum\limits_{m = 1}^M {{\rm{tr}}\left( {{{({\bf{F}}({{{\bf{\bar g}}}_m}))}^{ - 1}} + \frac{{{{\bf{C}}_m}}}{{{{{f}}}(\sigma_m ) + {c_m}}}} \right)}, ~~\label{eq:pemmse1}
\end{eqnarray}
where \[{{\bf{C}}_m} = {\left({{\bf{F}}({{{\bf{\bar g}}}_m})} \right)^{ - 1}}{\bf{f}}({{{\bf{\bar g}}}_m},\sigma_m ){\left( {{\bf{f}}({{{\bf{\bar g}}}_m},\sigma_m )} \right)^{\mathsf T}}{\left( {{\bf{F}}({{{\bf{\bar g}}}_m})} \right)^{ - 1}},\]
\[{c_m} = {\left( {{\bf{f}}({{{\bf{\bar g}}}_m},\sigma_m )} \right)^{\mathsf T}}{\left( {{\bf{F}}({{{\bf{\bar g}}}_m})} \right)^{ - 1}}{\bf{f}}({{{\bf{\bar g}}}_m},\sigma_m ) > 0.\]

Similarly, according to \eqref{eq:expectation} and  \eqref{eq:maxi}, the MSE of the EM algorithm is given by
\begin{eqnarray}
{\rm{ms}}{{\rm{e}}_{{\rm{EM}}}}({\bf{\hat H}})\mathop  = \limits^{{\rm{a}}{\rm{.e}}{\rm{.}}} {\rm{tr}}\left( {{{({\bf{F}}({\bf{H}}))}^{ - 1}} + \frac{{\bf{C}}}{{{f}(\sigma ) + c}}} \right),
\label{eq:emmse}
\end{eqnarray}
where ${\bf{F}}({\bf{H}})$ is block diagonal such that
\[{\bf{F}}({\bf{H}}) = \left[ {\begin{array}{*{20}{c}}
	{{\bf{F}}({{{\bf{\bar g}}}_1})}& \cdots &{\bf{0}}\\
	\vdots & \ddots & \vdots \\
	{\bf{0}}& \cdots &{{\bf{F}}({{{\bf{\bar g}}}_M})}
	\end{array}} \right],\]
\[{\bf{C}} = {\left( {{\bf{F}}({\bf{H}})} \right)^{ - 1}}{\bf{f}}({\bf{H}},\sigma ){\left( {{\bf{f}}({\bf{H}},\sigma )} \right)^{\mathsf T}}{\left( {{\bf{F}}({\bf{H}})} \right)^{ - 1}},\]
\[c = {({\bf{f}}({\bf{H}},\sigma ))^{\mathsf T}}{\left( {{\bf{F}}({\bf{H}})} \right)^{ - 1}}{\bf{f}}({\bf{H}},\sigma ),\]
\[{\bf{f}}({\bf{H}},\sigma ) = {[{({\bf{f}}({{{\bf{\bar g}}}_1},\sigma ))^{\mathsf T}}, {({\bf{f}}({{{\bf{\bar g}}}_2},\sigma ))^{\mathsf T}}, \cdots ,{({\bf{f}}({{{\bf{\bar g}}}_M},\sigma ))^{\mathsf T}}]^{\mathsf T}}.\]

Based on the block diagonal nature of the FIM, we have ${\mathop{ f}\nolimits} (\sigma ) = \sum\nolimits_{m = 1}^M {{{{f}}_m}(\sigma )}$, and $c = \sum\nolimits_{m = 1}^M {{c_m}}$. By subtracting \eqref{eq:pemmse1} from \eqref{eq:emmse}, we have
\begin{eqnarray}
d \!\!\!\!\!&=&\!\!\!\!\! {\rm{tr}} \left( {\sum\limits_{m = 1}^M {\frac{{{{\bf{C}}_m}}}{{{{{f}}_m}(\sigma ) + {c_m}}}}  - \frac{{\sum\nolimits_{m = 1}^M {{{\bf{C}}_m}} }}{{\sum\nolimits_{m = 1}^M {\left( {{{{f}}_m}(\sigma ) + {c_m}} \right)} }}} \right) \nonumber\\
\!\!\!\!\!&=&\!\!\!\!\! \sum\limits_{m = 1}^M {\left( {\frac{1}{{{{{f}}_m}(\sigma ) + {c_m}}} - \frac{1}{{\sum\nolimits_{n = 1}^M {\left( {{{{f}}_n}(\sigma ) + {c_n}} \right)} }}} \right)} {\rm{tr}}\left( {{{\bf{C}}_m}} \right) > 0,
\label{eq:d}
\end{eqnarray}
which means that the MSE of PEM is higher than that of EM.  Moreover, we observe that the terms ${c_m}$ and ${\rm{tr}}\left( {{{\bf{C}}_m}} \right)$ in \eqref{eq:d} decrease monotonically to zero while ${{{f}}_m}$ becomes a constant when the SNR approaches to zero. This implies that at the low SNR region, both EM and PEM yields an almost identical MSE, because $d \to 0$.

\end{document}